\documentclass[twocolumn,twocolappendix]{aastex631}
\usepackage{CJK}
\usepackage{multirow}

\usepackage{amsmath} 
\usepackage{comment}
\usepackage{bm}
\usepackage{xcolor} 
\newcommand{\SNnoFUVDT}{\texttt{SN-DT}}
\newcommand{\SNnoFUVML}{\texttt{SN-ML}}
\graphicspath{{./}{figures/}}

\begin{document}
\begin{CJK*}{UTF8}{min}

\title{ASURA-FDPS-ML: Star-by-star Galaxy Simulations Accelerated by Surrogate Modeling for Supernova Feedback}

\correspondingauthor{Keiya Hirashima}
\email{keiya.hirashima@riken.jp}

\author[0000-0002-1972-2674]{Keiya Hirashima (平島敬也)}\thanks{JSPS Research Fellow}
\affiliation{Department of Astronomy, Graduate School of Science, The University of Tokyo, 7-3-1 Hongo, Bunkyo-ku, Tokyo 113-0033, Japan}
\affiliation{Center for Computational Astrophysics, Flatiron Institute, 162 5th Avenue, New York, NY 10010, USA}
\affiliation{RIKEN Center for Interdisciplinary Theoretical and Mathematical Sciences (iTHEMS), RIKEN, Wako 351-0198, Japan}

\author[0000-0003-3349-4070]{Kana Moriwaki}
\affiliation{Research Center for the Early Universe, Graduate School of Science, The University of Tokyo, 7-3-1 Hongo, Bunkyo, Tokyo 113-0033, Japan}
\affiliation{Department of Physics, Graduate School of Science, The University of Tokyo, 7-3-1 Hongo, Bunkyo-ku, Tokyo 113-0033, Japan}

\author[0000-0002-6465-2978]{Michiko S. Fujii}
\affiliation{Department of Astronomy, Graduate School of Science, The University of Tokyo, 7-3-1 Hongo, Bunkyo-ku, Tokyo 113-0033, Japan}

\author[0000-0002-5661-033X]{Yutaka Hirai}
\affiliation{Department of Community Service and Science, Tohoku University of Community Service and Science, 3-5-1 Iimoriyama, Sakata, Yamagata 998-8580, Japan}

\author[0000-0001-8226-4592]{Takayuki R. Saitoh}
\affiliation{Department of Planetology, Graduate School of Science, Kobe University, 1-1 Rokkodai-cho, Nada-ku, Kobe, Hyogo 657-8501, Japan}
\affiliation{Center for Planetary Science (CPS), Graduate School of Science, Kobe
University 1-1 Rokkodai, Nada-ku, Kobe, Hyogo 657-8501, Japan}

\author[0000-0002-0411-4297]{Junichiro Makino}
\affiliation{Department of Planetology, Graduate School of Science, Kobe University, 1-1 Rokkodai-cho, Nada-ku, Kobe, Hyogo 657-8501, Japan}
\affiliation{Preferred Networks, Inc., 1-6-1 Otemachi, Chiyoda-ku, Tokyo 100-0004, Japan}
\affiliation{Center for Planetary Science (CPS), Graduate School of Science, Kobe
University 1-1 Rokkodai, Nada-ku, Kobe, Hyogo 657-8501, Japan}

\author[0000-0001-8867-5026]{Ulrich P. Steinwandel}
\affiliation{Center for Computational Astrophysics, Flatiron Institute, 162 5th Avenue, New York, NY 10010, USA}

\author{Shirley Ho}
\affiliation{Center for Computational Astrophysics, Flatiron Institute, 162 5th Avenue, New York, NY 10010, USA}
\affiliation{Department of Physics and Center for Data Science, New York University, New York, NY, USA}
\affiliation{Department of Astrophysical Sciences, Princeton University, Peyton Hall, Princeton, NJ 08544, USA}

\begin{abstract}
We introduce new high-resolution galaxy simulations accelerated by a surrogate model that reduces the computation cost by approximately 75 percent. Massive stars with a Zero Age Main Sequence mass of more than about 10 $\mathrm{M_\odot}$ explode as core-collapse supernovae (CCSNe), which play a critical role in galaxy formation. The energy released by CCSNe is essential for regulating star formation and driving feedback processes in the interstellar medium (ISM). 
However, the short integration timesteps required for SNe feedback have presented significant bottlenecks in astrophysical simulations across various scales.
Overcoming this challenge is crucial for enabling star-by-star galaxy simulations, which aim to capture the dynamics of individual stars and the inhomogeneous shell's expansion within the turbulent ISM.
To address this, our new framework combines direct numerical simulations and surrogate modeling, including machine learning and Gibbs sampling. The star formation history and the time evolution of outflow rates in the galaxy match those obtained from resolved direct numerical simulations. Our new approach achieves high-resolution fidelity while reducing computational costs, effectively bridging the physical scale gap and enabling multi-scale simulations.
\end{abstract}

\keywords{Galactic winds (572) --- Galaxy evolution (594) --- Hydrodynamical simulations (767) --- Stellar feedback (1602) --- Interstellar medium (847)}

\section{Introduction} \label{sec:intro}
Supernovae (SNe) are powerful events that release an immense amount of energy into the ambient ISM ($\sim 10^{51}$ erg). 
This energy release results in immediate heating and delayed momentum input into the ambient ISM, driving turbulence and aiding star formation regulation in the turbulent ISM \citep[e.g.,][for reviews]{Somerville+15, Naab+17}. 
While SNe strongly dominate the energy budget for galaxy formation and evolution, the evolution of galaxies also involves complex interactions among gravity, hydrodynamics, radiation, star formation, and chemical reactions. Given the complicated interplay among these processes, numerical methods have been commonly employed to study galaxy formation and evolution.

Recently, numerical simulations have reached the resolutions of $\sim 1 ~\mathrm{M_\odot}$ \citep[e.g.,][]{Andersson+2023, Andersson+2024, Deng+2024, Emerick+2018, Emerick+2019, Fujii+2021, Fujii+2021a, Fujii+2022, Fujii+2022a, Fujii+2024, Gutcke+2021, Gutcke+2022, Hirai+2021, Hirai+2025, Hu+2016, Hu+2017, Hu+2019, Lahen+2020, Lahen+2023, Smith+21, Steinwandel+2020, Steinwandel+23, Steinwandel+23a, Lahen+24, Steinwandel+24a, Steinwandel+24b}, which allow to resolve several important phases of SN-remnant evolution including the energy-conserving Sedov-Taylor phase, the pressure-driven snowplow phase, and the momentum-conserving snowplow phase \citep[e.g.,][]{Sedov1959, Taylor1950, Blondin+1998, vonNeumann1942} from first principles.

The timestep constraint of these simulations comes at a significant computational expense. Longer timesteps than the time-scale of cooling or adiabatic expansion tend to fail to estimate the internal energy of the gas and could lead to catastrophic results \citep[e.g.,][]{SaitohMakino09}. 
Capturing the Sedov phase, the initial evolution phase of explosions just after the energy injection, requires a timescale of hundreds of years, three orders of magnitude shorter than that of the typical ISM in galaxy simulations.
Proper timestep limiters are typically employed to preserve thermal energy injection by limiting the timestep with the signal velocity of a fluid tracer \citep{Monaghan1997, Durier+12, SaitohMakino09}. Even colder smoothed particle hydrodynamics (SPH) particles should have a smaller timestep to prevent vastly different time steps between neighboring SPH particles \citep[e.g.,][]{Chaikin+23, Nobels+23}. Hence, it might be desirable to speed up the injection scheme in a physically motivated manner.

Machine learning (ML) may help to extend the range of physical scales in computationally expensive simulations \citep[e.g.,][]{Kochkov+21}.
In cosmological simulations, ML has been used to predict the gravitational dynamics, allowing to avoid the computation of the physical states of $>10^9$ particles over the periods of the universe's age \citep[][]{He+19, Jamieson+23, Jamieson+24}.
ML has also been incorporated into hydrodynamical simulations to reconstruct the physical states of fluid tracer particles undergoing complex astrophysical interactions in realistic systems \citep[][]{Hirashima+23a, Hirashima+23b, Chan+24}.

Recent surrogate models for hydrodynamical simulations include two different approaches: physics-informed neural networks \citep[PINNs;][]{Raissi+19} and data-driven approaches.
PINNs explicitly use the governing equations for the loss function.
They can be used for both Eulerian-frame and Lagrangian-frame dataset \citep[e.g.,][]{Woodward+23}, so in principle, they can be applied to simulation data in a Lagrangian frame like galaxy formation simulation.
However, simulations of galaxy formation involve several equations for complex physics problems (gravity + hydrodynamics) that couple non-linear to other complex aspects of the astrophysical thermo chemistry and feedback physics, which in turn makes it quite hard to capture the physical state by analytic considerations alone and different model setups are needed to marginalize over all of the involved model parameters.
Furthermore, the rollout prediction in PINNs still requires small timesteps, which is a clear drawback for galaxy formation and evolution applications as we try to work around the dynamic range problem of astrophysical simulations. Nevertheless, there are successful PINNs for hydrodynamics under self-gravity, as demonstrated by \citet{Auddy+24}.

An alternative is the data-driven approach, which studies spatial and temporal correlations in the distribution of physical data without prior knowledge.
This approach has been applied for non-linear and multiscale physical simulations in astrophysics in recent work \citep{He+19, Bernardini+22, Hirashima+23a, Hirashima+23b, Jamieson+23, Jamieson+24, Chan+24, Legin+24}. In this paper, instead of PINNs, we present the first data-driven surrogate modeling using ML to accelerate galaxy formation simulations by mitigating the bottlenecks caused by small timesteps.

The paper is structured as follows. In Sec.~\ref{sec:style}, we discuss the numerical methods used. In Sec.~\ref{sec:SN}, we test our network prediction against a resolved SN-blastwave in a turbulent molecular cloud. In Sec.~\ref{sec:dwarf}, we apply this network in a resolved dwarf galaxy simulation with explicit, star-by-star stellar feedback and compare our network results to direct numerical simulations with thousands of individually resolved SN-explosions and compare the SN-environmental density distributions and the outflow rates. In Sec.~\ref{sec:discussion}, we discuss our results in the context of previous star-by-star simulations of dwarf galaxies with an explicitly resolved ISM. In Sec.~\ref{sec:conclusions}, we conclude and summarize our results.

\section{Methods} \label{sec:style}
\subsection{Simulation code \label{sec:sim_code}}
This study executes three types of runs: high-resolution simulations of SN feedback in isolated molecular clouds, a fiducial run of an isolated dwarf galaxy, and a run of the same galaxy enhanced using the surrogate model for SN feedback.
The simulation data of isolated molecular clouds are used to train the surrogate model.
The isolated SN simulations and the fiducial run of the galaxy are carried out with our $N$-body/density-independent smoothed particle hydrodynamics \citep[DISPH;][]{Saitoh+2013, Saitoh+2016} code, \textsc{ASURA-FDPS} \citep{Saitoh+2008,Iwasawa+2016, Hirashima+23a}. Hydrodynamic interactions are computed by the DISPH method \citep[][]{Saitoh+2013, Saitoh+2016}, which can properly handle the contact discontinuity.
The gas properties are smoothed using the Wendland $C^4$ kernel \citep[][]{Dehnen+12}.
The kernel size is determined to maintain the number of neighbor particles at $125\pm1$.
We adopt the shared variable timesteps with the Courant-like hydrodynamical timestep \citep{Courant+1928} based on the signal velocity \citep[][]{Monaghan1997, Springel2005}. 
Following the leapfrog (Velocity Verlet) scheme, we adopt a second-order symplectic integrator in its Kick-Drift-Kick (KDK) configuration. In the simulations of SN feedback with a high-mass resolution, the timestep of a SPH particle $i$ is determined as
\begin{equation}
    \Delta t_i = C_{\rm CFL} \frac{2 h_i}{\max_{j} [c_i + c_j - 3w_{ij}]}
    \label{eq:CFL}
\end{equation}
where $C_{\rm CFL}=0.3$, $h_i$ and $c_i$ are the SPH smoothing length and sound speed of a particle $i$, and 
{\bf
\begin{equation}
w_{ij} =
\begin{cases} 
    {\bm v}_{ij} \cdot {\bm r}_{ij}/r_{ij} & ( {\bm v}_{ij} \cdot {\bm r}_{ij} < 0 ), \\
    0, &  ( {\bm v}_{ij} \cdot {\bm r}_{ij} \geq 0 ).
\end{cases}
\end{equation}
}
where ${\bm r}_{ij}$ and ${\bm v}_{ij}$ are the relative position and velocity between particles $i$ and $j$, respectively.

Our new presenting framework \textsc{ASURA-FDPS-ML} consists of \textsc{ASURA-FDPS} and the surrogate modeling for SN feedback, which will be described in Sec. \ref{sec:framework}.
In our dwarf galaxy simulations with surrogate modeling, we fix timesteps to 2000 years by avoiding direct calculations of the initial state of SN feedback, which significantly requires smaller timesteps.

\subsection{Initial Condition}
The initial condition (IC) for the isolated galaxy simulations are originally described in \citet{Hu+2016,Hu+2017} (``cmp"). This IC has been used to investigate the properties of the dwarf galaxy and treatment of stellar feedback \citep[e.g.,][]{Smith+21, Hislop+2022, Steinwandel+23, Hu+23}. The ICs were generated with the method proposed in \citet{Springel2005}. The dark matter halo follows a Hernquist profile with an NFW-equivalent \citep{Navarro+97} concentration parameter $c_\mathrm{NFW} =10$ and has a virial radius of $R_\mathrm{vir}=44$ kpc and a viral mass of $M_\mathrm{vir} = 2 \times 10^{10} ~\mathrm{M_\odot}$. The initial gas mass of the system is $4 \times 10^7 ~\mathrm{M_\odot}$, and the stellar background potential has a mass of $2 \times 10^7 ~\mathrm{M_\odot}$.
The initial disk consists of 4 million dark matter particles, 10 million gas particles, and 5 million star particles, setting a dark matter particle mass resolution of $m_\mathrm{DM} = 6.8 \times 10^3 ~\mathrm{M_\odot}$ and a baryonic particle mass resolution of $m_\mathrm{baryon} = 4 ~\mathrm{M_\odot}$. The gravitational softening parameter of the simulations is set to 62 pc for dark matter and 0.5 pc for gas and stars. Unlike some previous studies, the IC used in this paper has evolved in advance for 500 Myr with the code \textsc{P-Gadget3} in the MFM version of \citet{Steinwandel+2020} with 32 neighbors for the density computation. 
Fluid fluxes up to that point have been computed with the Harten-Lax-van-Leer contact (HLLC) Riemann solver. After the initial 500 Myr period, we completely switch to the \textsc{ASURA-FDPS} (or \textsc{ASURA-FDPS-ML}) framework based on the DISPH method outlined in Sec.~\ref{sec:sim_code}. 
We note that only a small amount of gas is converted into stars during the pre-evolution phase, and, for simplicity, we ignore them in the subsequent evolution. See Appendix \ref{app:pre-evolution} for further discussion on the switching effects.
We adopt an initial metallicity of 0.1 $\mathrm{Z_\odot}$ following the solar abundance pattern and only use the stellar background particles, excluding massive stars, when switching the framework.

\subsection{Star formation model}
Gas particles stochastically form stars if they satisfy several physical conditions such as high number density of hydrogen atoms ($n_\mathrm{H} \geq 100 ~\mathrm{cm^{-3}}$), low temperature ($T \leq 100$ K), and convergence of the flows ($\nabla \cdot v < 0$), following \citet{Hirai+2021}.
To avoid star formation in hot regions, which have a large Jeans mass, the threshold temperature 100 K is chosen following \citet{Hu+2016}.
The masses are sampled by the initial mass function (IMF), following \citet{Chabrier+03} with an upper limit of 40 M$_\odot$. The lower limit for the IMF is 0.1 M$_\odot$.
When a gas particle meets the physical conditions, it is converted into a star with a probability ($p_*$) in a given timestep ($\Delta t$):
\begin{equation}
    p_* = \left \{ 1 - \exp \left( -c_* \frac{m_\mathrm{gas}}{\langle m_* \rangle}  \frac{\Delta t}{t_\mathrm{dyn}} \right) \right \},
\end{equation}
where $m_\mathrm{gas}$, $\langle m_* \rangle$, $t_\mathrm{dyn} = 1/\sqrt{4\pi G \rho}$, and $c_*$ are the mass of one gas particle, the average value of stellar mass in the assumed IMF, the local dynamical time of the star-forming region, and a fixed dimensionless star-formation efficiency as 0.02 per dynamical time \citep{Krumholz+19}.

\subsection{Stellar feedback}
We used the yields from \citet{Nomoto+13}, which covers core-collapse SNe from 13 M$_\odot$ to 40 M$_\odot$, where the nucleosynthesis yields as masses of individual elements are given as functions of the stellar mass, metallicity, and explosion energy of 10$^{51}$ erg. The lifetimes are determined depending on the metallicity using the table in \citet[][]{Portinari+98}. These are compiled in the Chemical Evolution Library \citep[\textsc{CELib},][]{Saitoh+2017}.

\subsection{Required timestep for SN feedback}
\label{sec:rec_timestep}
The timestep in \textsc{ASURA-FDPS-ML} is fixed with a reasonably short time for a simple implementation described in Section \ref{sec:framework}.
We estimate the required timestep using a time scale for discrete SN feedback shown in \citet{SaitohMakino2010}.
Here, the size of an SPH particle, $\lambda$, is assumed to be
\begin{equation}
    \lambda = \left(\frac{3}{4\pi} \frac{m_\mathrm{SPH}}{\rho}\right)^{1/3},
    \label{eq:size}
\end{equation}
where $\rho$ is the density of a SPH particle.
Let $N_\mathrm{NB}$ be the number of neighboring particles that receive energy from a SN. The internal energy for a SPH particle due to the SN of a single star, $U_{\mathrm{SN}}$, is defined as
\begin{align}
    U_{\mathrm{SN}} &:= \frac{E_{\mathrm{SN}}}{N_\mathrm{NB} ~m_\mathrm{SPH}} \nonumber\\
    & = \frac{10^{51}}{N_\mathrm{NB} ~m_\mathrm{SPH}} ~[\mathrm{erg~M_\odot^{-1}}] \nonumber\\
    & \simeq \frac{5.0 \times 10^{17}}{N_\mathrm{NB}} ~ \left( \frac{1~ \mathrm{M_\odot}}{m_\mathrm{SPH}}\right) ~[\mathrm{erg~g^{-1}}],
    \label{eq:USN}
\end{align}
where the original internal energy is ignored since it is low in comparison to the SN-ejecta.
With Equation (\ref{eq:USN}), the sound speed, $c_{\mathrm{SN}}$, of the heated gas is derived as
\begin{align}
    c_{\mathrm{SN}} &:= \sqrt{\gamma(\gamma-1)U_{\mathrm{SN}}} \nonumber\\
    & \simeq \frac{7.5 \times 10^{3}}{N_\mathrm{NB}^{1/2}} ~ \left( \frac{1~ \mathrm{M_\odot}}{m_\mathrm{SPH}}\right)^{1/2} ~[\mathrm{km~s^{-1}}].
    \label{eq:cSN}
\end{align}
By substituting Equations (\ref{eq:size}) and (\ref{eq:cSN}) and $N_\mathrm{NB}=128$ into the sound crossing time, $t_\mathrm{SN} \equiv \lambda / c_\mathrm{SN}$,
\begin{equation}
    t_\mathrm{SN} \simeq 1.3 \times 10^3 \left( \frac{m_\mathrm{SPH}}{1~ \mathrm{M_\odot}}\right)^{5/6} \left( \frac{100 ~[\mathrm{cm^{-3}}]}{n_\mathrm{H} }\right)^{1/3} ~[\mathrm{yr}].
    \label{eq:sound_crossing}
\end{equation}
To avoid overcooling of SN feedback in star-forming regions, we replace the direct computation of SN feedback in denser regions than a hydrogen number density of $n_\mathrm{H}=1~\mathrm{cm^{-3}}$ with our surrogate model including ML.
Using Equation (\ref{eq:sound_crossing}), the timescale for SN feedback is calculated as $t_\mathrm{SN} \simeq 2 \times 10^4$ years at a mean density of $n_\mathrm{H}=1~\mathrm{cm^{-3}}$ when the mass resolution $m_\mathrm{SPH}=4~M_\odot$ is applied.
Given that the typical Courant time-step in the region is one-tenth of the timescale of SNe, we set a fixed timestep $\Delta t = 0.1\,t_\mathrm{SN} = 2 \times 10^3$ years, sufficiently small to resolve SNe in the ISM with a density lower than $1~\mathrm{cm^{-3}}$.

We also note that, in simulations with a fixed timestep in our new framework, the Courant condition of equation (\ref{eq:CFL}) is satisfied for almost all particles.
On average, only 0.08 particles (0.32 $\mathrm{M_\odot}$) per snapshot require timesteps shorter than the fixed timestep of 2000. Most of these particles require approximately 1900 years, with the shortest requiring about 1400 years. This difference is less than 30 percent smaller than the fixed timestep, and since these shorter timesteps are conservatively estimated with the relative velocity as $3w_{ij}$ using equation (\ref{eq:CFL}), such minor differences are not significant. Therefore, we conclude that while a tiny number of particles technically violate the Courant condition, their impact on the simulation results is negligible due to both their small number and the minor degree of violation.

\subsection{Hybrid on-the-fly ML-driven simulations}
\label{sec:framework}
\begin{figure*}[ht!]
\includegraphics[width=\textwidth]{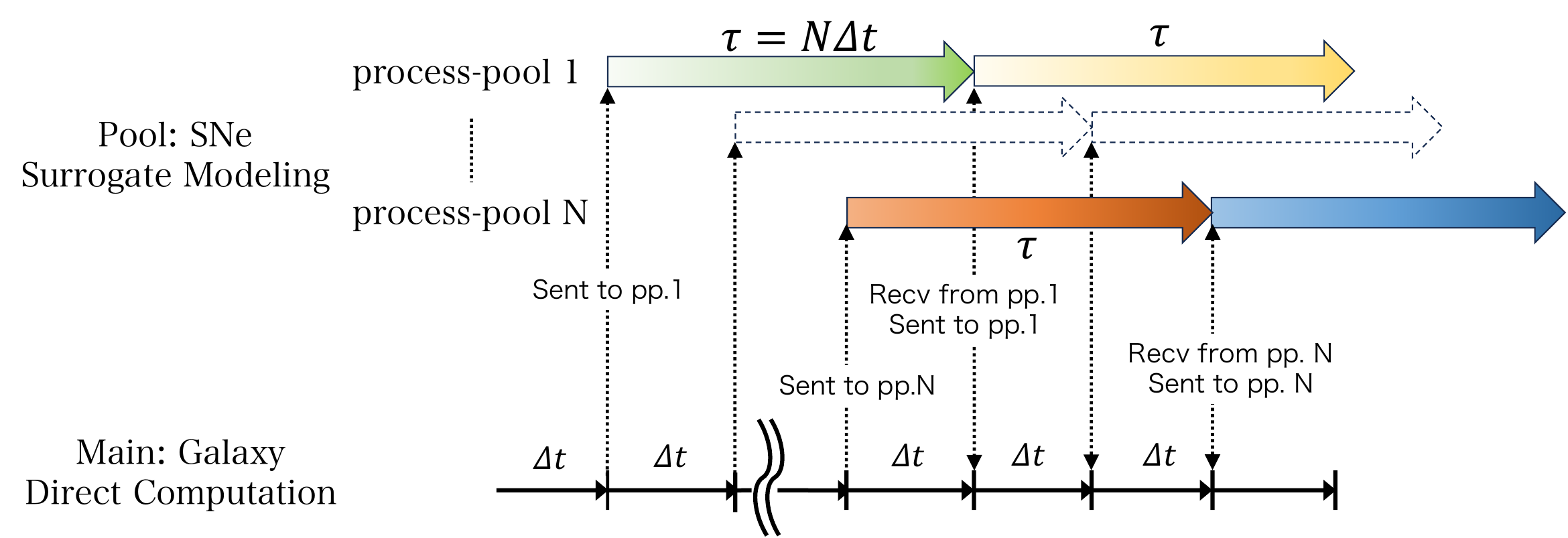}
\caption{Schematic diagram for our presenting hybrid on-the-fly ML-driven framework.
\label{fig:bridge}}
\end{figure*}

Figure \ref{fig:bridge} shows the schematic diagram of our presenting framework, \textsc{ASURA-FDPS-ML}. It has two kinds of MPI communicators: ``Main'' and ``Pool'' hereafter. 
A whole galaxy, including $N$-body and SPH, is directly computed in main processes using hundreds of MPI processes (``Main''), while SN feedback in dense regions ($>1~\mathrm{cm^{-3}}$) is handled by a surrogate model in ``Pool'' processes. 
As SNe in the less dense ambient gas are more kinetically resolved \citep{Kim+2015,Steinwandel+2020,Gutcke+2021}, when SNe explode in the environments at a hydrogen number density of $< 1 \mathrm{cm^{-3}}$, the energy of SNe ($10^{51}$ erg) is injected into the neighboring 100 SPH particles weighted using the Wendland $C^4$ kernel \citep[][]{Dehnen+12}.
Every timestep, ``Main'' checks if any SN occurs in dense regions ($>1~\mathrm{cm^{-3}}$) or not. If SNe occur, the ambient gas particles are sent to a ``Pool'' process.
The surrogate model in the ``Pool'' process reconstructs the distribution of gas particles after a time window $\tau$ from the SN explosion. Once the ``Main'' processes are advanced for $\tau$, the predicted gas particles are returned to ``Main'' from ``Pool''.
If multiple SNe occur in a dense, compact region, our surrogate model handles the first explosion, which reduces the local gas density. Subsequent SNe then occur in a lower-density environment ($< 1~\mathrm{cm^{-3}}$) and are automatically treated by the direct calculation method.

To prevent the evolution in the replaced regions from being violated by gravity, we restrict the time window using the local free-fall time $t_\mathrm{ff}$ at SF regions, where the density could potentially reach $\sim 10^3~\mathrm{cm^{-3}}$ in the simulations.
As the typical gravitational timestep is the one-tenth of the local free-fall time, by using $t_\mathrm{ff}\sim 1$ Myr at $n_\mathrm{H}=10^3~\mathrm{cm^{-3}}$, we set $\tau = 0.1\,t_\mathrm{ff} = 0.1$ Myr\footnote{This is also the timescale on which the SN remnant transits from the energy conserving phase to the momentum conserving phase (see Figure \ref{fig:conv_SPH}).}.
By this definition, the number of MPI processes in ``Pool'' is set to $N=\tau / \Delta t=50$ while ``Main'' can have an arbitrary number of MPI processes.

\subsection{Dataset for Our Surrogate Modeling} \label{sec:ddsurogate}
To avoid direct computation of the regions that require small timesteps, i.e., the SN explosions at a hydrogen number density of $> 1~\mathrm{cm^{-3}}$, we used the simulations of SN feedback in such environments as the training dataset for our ML model \citep{Hirashima+23a,Hirashima+23b}.
The simulations are designed as a SN explosion in a high-density star-forming molecular cloud with a large density contrast.
We assume an adiabatic compression of a monoatomic ideal gas, which follows the equation of state with the specific heat ratio $\gamma=5/3$:
\begin{equation}
    P=(\gamma-1) \rho u,
\end{equation}
where $P$, $\rho$, and $u$ are the pressure, smoothed density, and specific internal energy, respectively.
The adiabatic compressible gas clouds follow the following equations:
\begin{align}
    \frac{d \rho}{dt} & = -\rho \nabla \cdot \bm{v}, \\
    \frac{d^2 \bm{r}}{dt^2} & = -\frac{\nabla P}{\rho} + \bm{a}_{\rm visc}-\nabla \Phi, \\
    \frac{d u}{dt} & = -\frac{P}{\rho} \nabla \cdot \bm{v} + \frac{\Gamma-\Lambda}{\rho},
\end{align}
where $r$ is the position, $a_{\rm visc}$ is the acceleration generated by the viscosity, $\Phi$ is the gravitational potential, $\Gamma$ is the radiative cooling rate per unit volume, and $\Lambda$ is the radiative heating rate per unit volume.

Initially, uniform and virialized gas spheres with a homogeneous isotropic turbulent velocity field that follows $\propto v^{-4}$ are constructed using the \textsc{Astrophysical Multi-purpose Software Environment} \citep[\textsc{AMUSE};][]{Pelupessy+2013, Portegies+2013, Portegies+2018}.
These initial gas spheres have a total mass of $10^6 ~\mathrm{M_\odot}$ and a radius of ~60 pc. All of the simulations are performed with a mass resolution of 1 $\mathrm{M_\odot}$, which is a higher resolution than the target resolution of 4 $\mathrm{M_\odot}$ for one of our galaxy simulations in this paper.
The gas sphere is evolved for one initial free-fall time $t_\mathrm{ff} = \left( \frac{3\pi}{32G\rho_0} \right)^{1/2}$ to mimic star-forming regions and supports the build-up of filamentary structures by decaying (supersonic) turbulence.
We inject the thermal energy of $10^{51}$ erg into 100 neighboring SPH particles of the center of mass of the turbulent gas clouds. The pairs of the snapshots at $t=0$ and $t=\tau =0.1$ Myr are used as the input and output, respectively.
Our dataset includes 400 simulations using 300 for training and 100 for testing.

\section{Test for our surrogate model of SN feedback} \label{sec:SN}

\begin{figure*}[ht!]
    \includegraphics[width=\textwidth]{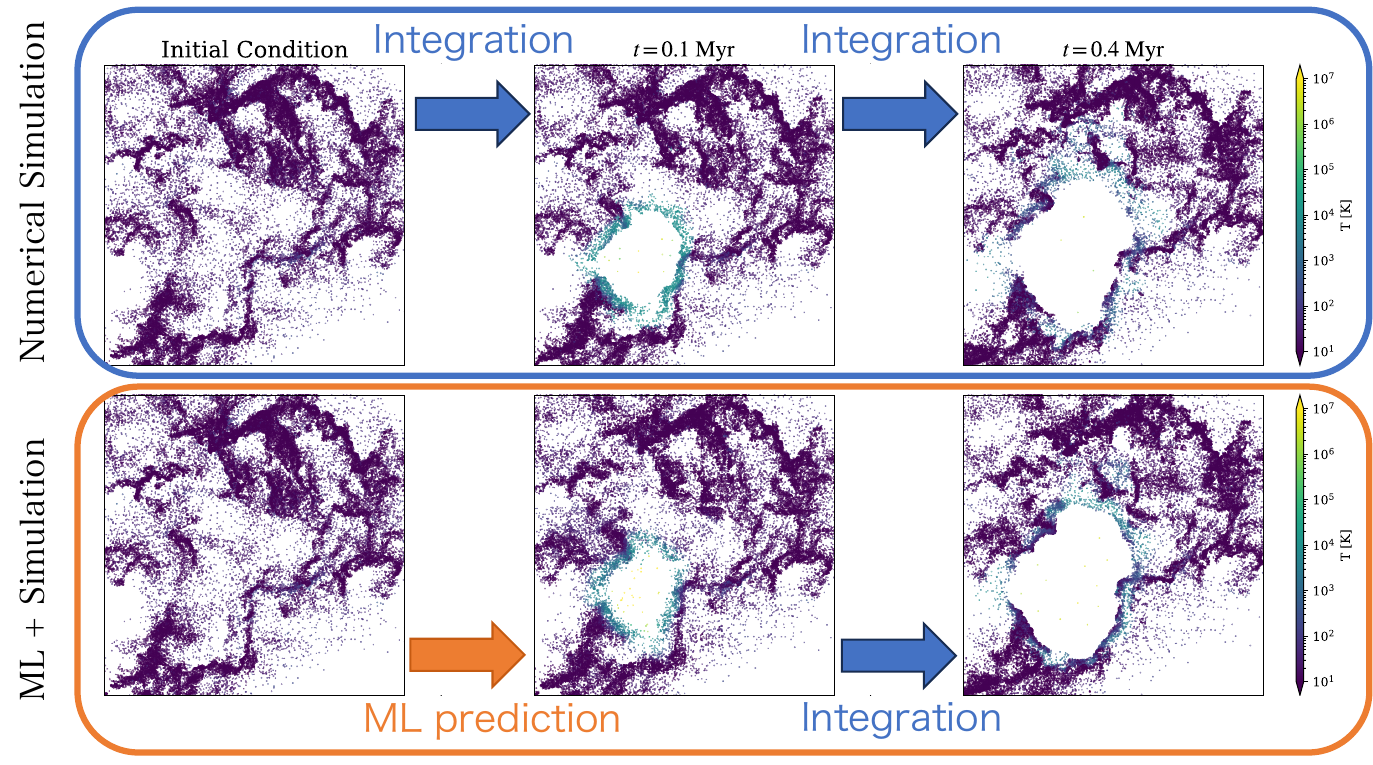}
    \caption{Comparison between simulations and ML for {\bf an} isolated SN. {\it Upper}: The numerical simulation results. {\it Lower}: The numerical simulation results with our surrogate SN feedback model.
    Cross section of snapshots at $t=0, ~10^5, ~ \mathrm{and} ~4 \times 10^5$ years are listed up from left to right.}
    \label{fig:CompSN}
\end{figure*}

Our surrogate model is designed to incorporate a ML model and Gibbs sampling and reconstruct the distribution of gas particles with temperature and 3D velocity.
Our ML model is based on U-Net \citep{Ronneberger+15}, which is a supervised ML model originally developed for computer vision and consists of convolutional neural networks. The model learns gas dynamics of SN feedback in 3D from simulations.
As described in Section \ref{sec:ddsurogate}, the ML model was trained with 300 simulations of an isolated SN in a molecular cloud with a total mass of $10^6$ $\mathrm{M_\odot}$ with a mass resolution of 1 $\mathrm{M_\odot}$ \citep{Hirashima+23b}.
The ML model takes the 3D distribution ($64^3$ voxels with one side of 60 pc and a spatial resolution of $\sim 1$ pc) of density, temperature, and 3D velocities as an input and predicts those fields after the SN explosion by the time window $\tau=0.1$ Myr as an output, completed in Section \ref{sec:framework}.
Since the time window is short enough compared to the local free-fall time, we assume the cold gas particles in the voxels barely move, and the number of particles (the total mass) inside is preserved over the time window $\tau$.
The model $\mathcal{M}$ is trained to learn the relation between input $X$ and output $y$, where $X$ and $y$ represent the distribution of physical quantities before and after the SN explosion in simulations, respectively, over the time window of 0.1 Myr.
Suppose the trainable parameter $\theta$ and the predicted distribution $\hat{y}$, we summarize the procedure of the surrogate model as the following:
\begin{enumerate}
    \item Make the input voxel $X$
    \item Predict physical distribution 0.1 Myr after the explosion $\hat{y} = \mathcal{M}(X \mid \theta)$
    \item Sample as many particles as the input
\end{enumerate}
Further details of our surrogate modeling are provided in Appendix \ref{sec:surrogate_model}.

Figure~\ref{fig:CompSN} compares a direct simulation (the upper panels) by \textsc{ASURA-FDPS} and a simulation incorporating our surrogate model for SN feedback (the lower panel) by \textsc{ASURA-FDPS-ML}.
The surrogate model reconstructs the asymmetric shell and hot region. 
The shell can evolve smoothly and stably even after the predicted particles return to the ``Main'' calculation domain.
We also show the conversation of thermal energy and outer momentum in the reconstruction using our surrogate model, comparing it to a low-resolution simulation in Appendix \ref{sec:fidelity}.

\section{Results for an isolated dwarf galaxy simulation} \label{sec:dwarf}
In this section, we present the results of a simulation of an isolated dwarf galaxy using the surrogate model for SN feedback.
The surrogate model is applied for only SNe in environments denser than 1 $\mathrm{cm^{-3}}$ as discussed in Section \ref{sec:framework}.
When SNe explode in the environments at a hydrogen number density of $< 1 \mathrm{cm^{-3}}$, the energy of SNe ($10^{51}$ erg) is injected into the neighboring 100 SPH particles weighted using the Wendland $C^4$ kernel \citep[][]{Dehnen+12}.

\begin{table*}
    \centering
    \caption{List of physical models of isolated galaxy simulations in this paper. }
    \begin{tabular}{lccl}
    \hline\hline
    Run Name & Supernova Feedback  &Time-Step& Note\\
    \hline
         \SNnoFUVDT & Thermal Feedback    &           Determined by eq. (\ref{eq:CFL})     &  Fiducial run\\
         \multirow{2}{*}{\SNnoFUVML} &  Thermal Feedback (if $n_\mathrm{H}< 1 \mathrm{cm^{-3}}$)  & \multirow{2}{*}{Fixed 2,000 yr}  & \\ 
                    &  Our Surrote Modeling  (if $n_\mathrm{H}> 1 \mathrm{cm^{-3}}$) &  &    \\  
         \hline
    \end{tabular}
    \label{tab:models}
\end{table*}

\subsection{Morphology Comparison}
\label{sec:morpho}
\begin{figure*}[ht!]
\begin{interactive}{animation}{Fig3_Column_density.mp4}
\includegraphics[width=\textwidth]{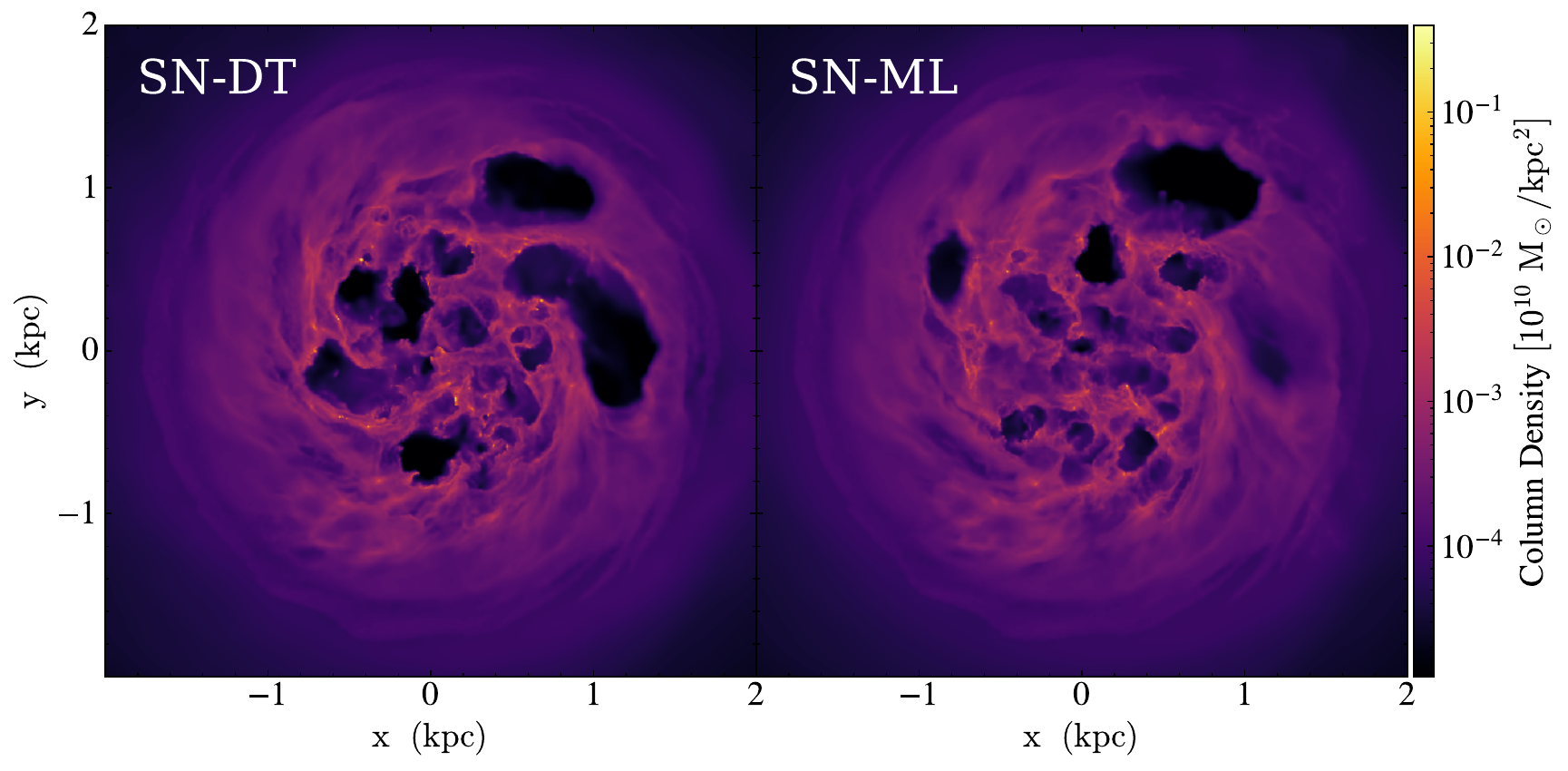}
\end{interactive}
\caption{
  Face-on surface density of gas at $t=100$ Myr for all models. Color represents column density in $\mathrm{10^{10}~M_\odot~kpc^{-2}}$. {\it Left:} \SNnoFUVDT. {\it Right:} \SNnoFUVML. The different model variations shown in this Figure are summarized in Table \ref{tab:models}.
  An animation showing the time evolution of the face-on gas surface density is available. It spans 300 Myr of simulation time and lasts 20 seconds in real time.
}
\label{fig:numGal}
\end{figure*}

Figure~\ref{fig:numGal} shows the face-on surface density of the gas for two models summarized in Table \ref{tab:models}, the fiducial run ({\SNnoFUVDT}) on the left and the run with ML ({\SNnoFUVML}) on the right. 
We find little difference between the two runs when it comes to the morphological structure of the galactic disc.
While the morphological structures that emerge on both simulations are super bubbles driven by clustered feedback, we note that these volume-filling structures are not identical in the two reference runs we carried out. The reason for this is twofold. First of all, the binary executables of the code are not identical, which will lead to small-scale differences in execution. Second, some of the physics modules, such as the IMF sampling for star formation, introduce some randomness in the exact stellar population that is obtained, which in turn influences the feedback cycle due to the different lifetimes of massive stars of a given mass.

The ISM structures in our simulations are visually similar to other recent simulations of similar isolated dwarf galaxies \citep[e.g.,][]{Smith+21,Deng+2024,Gutcke+2021,Hu+2019}.
Fine mass and time resolutions are generally required for accurately resolving such superbubbles; otherwise, the hot gas within the bubble cools too quickly, a problem typically referred to as numerical over-cooling  \citep[e.g.,][]{Pearce+99,Thacker+00b,Croft+01,Springel+02}.
In our new scheme \textsc{ASURA-FDPS-ML}, however, despite the longer timesteps, several distinct superbubbles emerged.

\subsection{Star Formation History}
\begin{figure}[ht!]
    \includegraphics[width=0.49\textwidth]{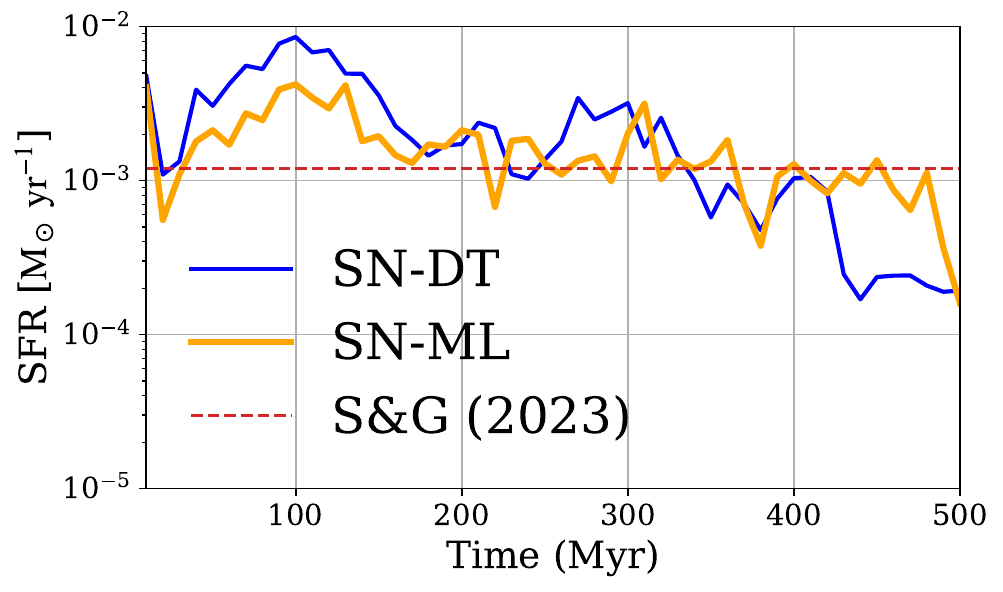}
    \caption{Star formation histories averaged with an interval of 10 Myr as a function of time for each run as listed in Table \ref{tab:models}. The red dashed line shows the fiducial run in \cite{Steinwandel+23}.}
    \label{fig:SFR}
\end{figure}

Figure~\ref{fig:SFR} shows the star formation history (SFH), representing the SFRs averaged with an interval of {\bf 10} Myr as a function of time.
For reference, the red dashed line shows the SFR obtained from the MFM simulation in \citet{Steinwandel+23} as in their WLM-fid simulation.
Since the IC of a disk dwarf galaxy does not have steller feedback initially, the galaxies have a high SFR from 0 to 80 Myr. 
The simulations, \SNnoFUVDT~and \SNnoFUVML, have a similar SFR trend, they marginally differ; \SNnoFUVDT~shows an excess of star formation activity in the early stages of evolution, while \SNnoFUVML~shows an excess of star formation activity in the late stages of the evolution. However, the agreement in the integrated star formation history (total stellar mass formed) is within a few percent between the two simulations, which is within so-called run-to-run variations that have been reported in previous studies of similar simulations \citep[e.g.,][]{Steinwandel+23a}.

\subsection{Multi-phase structures in ISM}
\begin{figure*}[ht!]
    \plotone{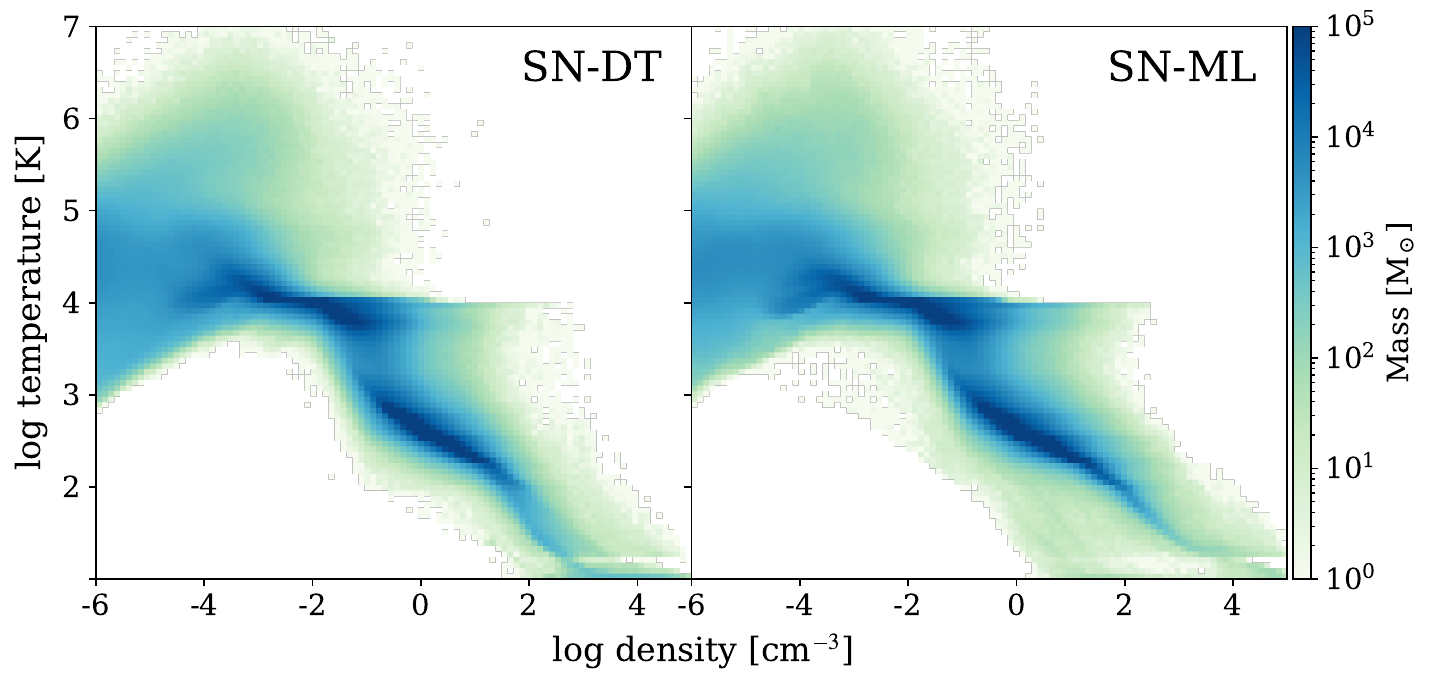}
    \caption{Phase structures of \SNnoFUVDT~({\it left}) and \SNnoFUVML~({\it right}) as listed in Table \ref{tab:models}. Density-temperature phase diagrams of mass are averaged between 300 Myr with an interval of 20 Myr.}
    \label{fig:PhaseSpace}
\end{figure*}

Figure~\ref{fig:PhaseSpace} shows the gas phase structures for the model \SNnoFUVDT~on the left and \SNnoFUVML~on the right. This is the two-dimensional distribution of mass in the density-temperature plane.
These are averaged over 300 Myr with a spacing of 20 Myr (i.e., over 15 snapshots). 
The two simulations agree remarkably well on the obtained ISM phase structure, having a three-phase SN-driven medium with a diffuse hot phase at low densities, a stable warm neutral (WNM) phase at intermediate densities, and a stable cold phase at high densities. Visually, the model \SNnoFUVML~has a more extended unstable ISM phase between 0.01 and 100 cm$^{-3}$, but we note that the difference in absolute mass is negligible. 

\subsection{Definition of Outflow Rates and Loading Factors}
We follow the definition of the discrete outflow rates for mass and energy proposed in literature \citep{Hu+2017,Hu+2019, Steinwandel+23} with the net total flow rates defined as $\dot{M}=\dot{M}_\mathrm{out} - \dot{M}_\mathrm{in}$ and $\dot{E}=\dot{E}_\mathrm{out} - \dot{E}_\mathrm{in}$.
Outflows are computed as the sum of all gas with a positive or negative vertical velocity ($v_z$) in the slab located above or below the disc, respectively.
Mass and energy outflow rates are defined as:
\begin{equation}
    \dot{M}_\mathrm{out} = \sum_{i, v_{z,i}>0} \frac{m_i v_{z,i}}{\Delta z},
\end{equation}
\begin{equation}
    \dot{E}_\mathrm{out} = \sum_{i, v_{z,i}>0} \frac{m_i( v_{i}^2 + \gamma u_i)v_{z,i}}{\Delta z},
\end{equation}
where $m_i$ and $u_i$ are the mass and specific internal energy of SPH gas element, and $\gamma = 5/3$.
We measure the outflow rates and loading factors at two different heights of 1 kpc and 10 kpc in slabs with radii of 5 kpc and thicknesses $\Delta z$ of 0.1 kpc and 1 kpc, respectively.
Normalizing these outflows by reference quantities such as the global averaged SFR and SN rate can be useful. Such a normalized quantity is called the loading factor.
\begin{equation}
    \eta_{m}^\mathrm{out} = \dot{M}_\mathrm{out} / \overline{\mathrm{SFR}}
\end{equation}
\begin{equation}
    \eta_{e}^\mathrm{out} = \dot{E}_\mathrm{out} / (E_\mathrm{SN} \overline{\mathcal{R}_\mathrm{SN}}),
\end{equation}
where we adopt $E_\mathrm{SN}=10^{51}$ erg. The mean star formation rate $\overline{\mathrm{SFR}}$ and the mean SN rate $\overline{\mathcal{R}_\mathrm{SN}}$ are derived from the simulations.
We also define Bernoulli velocity as
\begin{equation}
    v_B = \left( v_{z}^2 + \frac{2 \gamma}{\gamma -1} c_s^2 \right)^{1/2},
    \label{eq:Bernoulli}
\end{equation}
where $c_s := \sqrt{\gamma (\gamma -1 ) u}$ is the sound speed. 
The Bernoulli velocity can be interpreted as the terminal velocity that an adiabatic wind would reach under the assumption of $v_b \gg v_\mathrm{esc}$ where $v_\mathrm{esc}$ is an escape velocity.

We note that we find generally very good agreement with other studies in this field \citep{Hu+2019, Gutcke+2021, Hislop+2022, Steinwandel+23a, Steinwandel+23, Steinwandel+24a, Porter+24} and we will discuss this in greater detail in Section~\ref{sec:discussion}.

\subsection{Time Evolution and Phase Structure of the Outflows}

\begin{figure*}            
    \includegraphics[width=\textwidth]{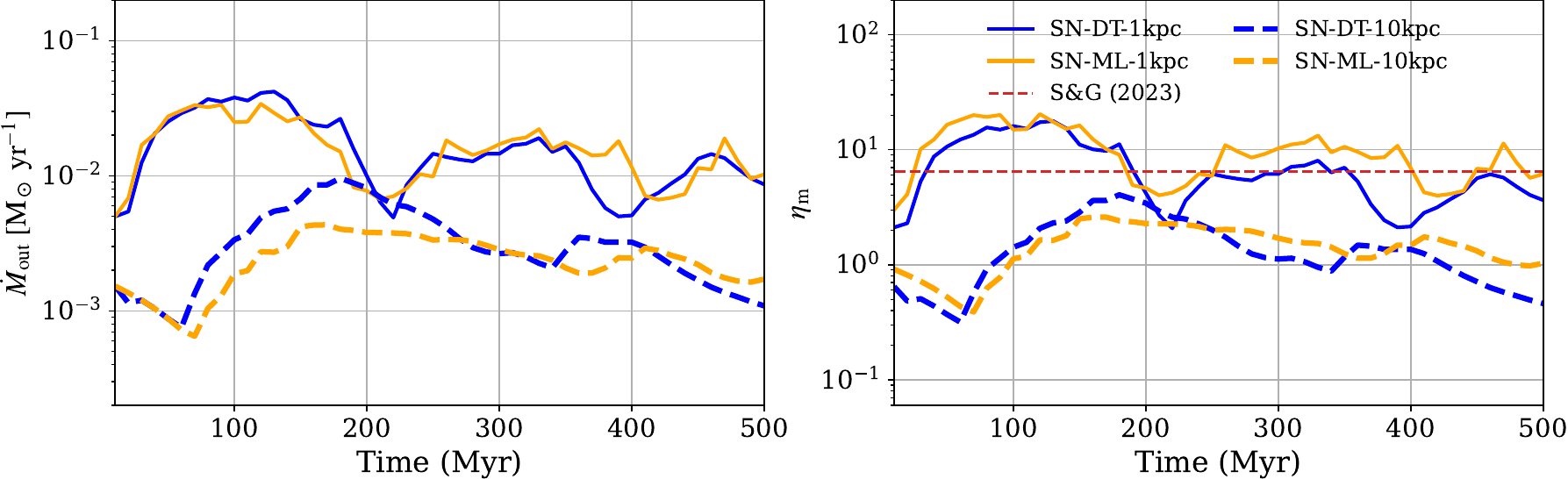}
    \includegraphics[width=\textwidth]{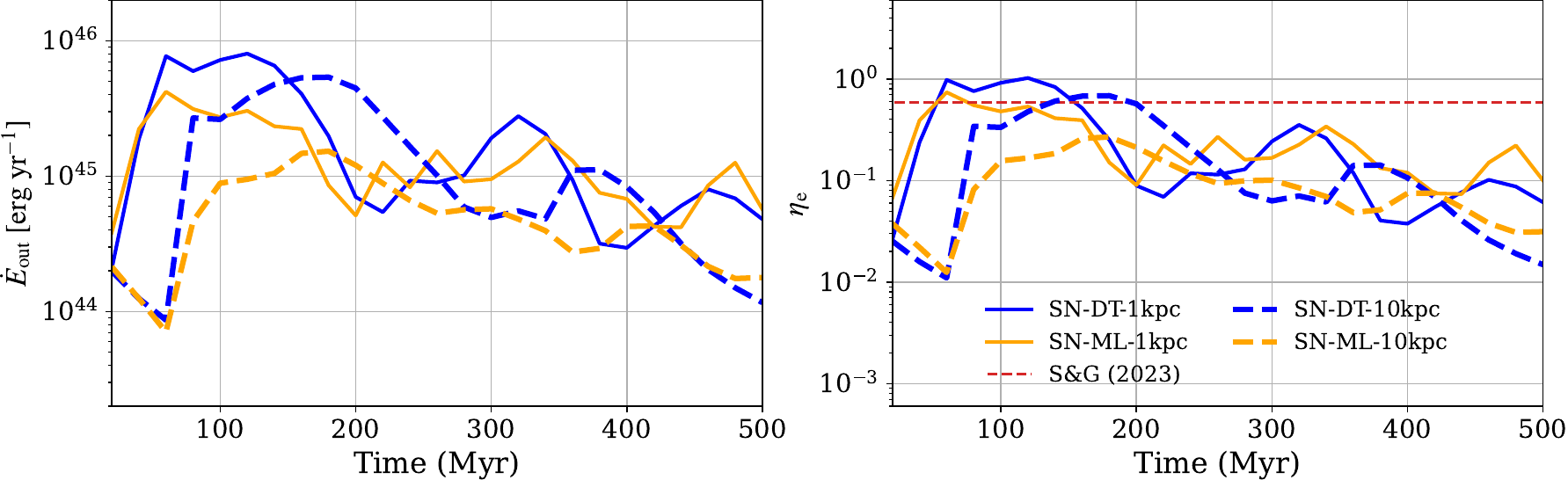}
    \caption{Outflows as a function of time for each run as listed in Table \ref{tab:models}. The Red dashed line shows the loading factors of the fiducial run in \cite{Steinwandel+23}. {\it Upper left}: Mass outflow rates of gas, which are the mass of the gas that crosses the slab with a thickness of 100 pc at a height of 1 kpc per year. {\it Upper right}: Mass loading factors for all runs as a function of time. {\it Lower left}: Energy outflow rates of gas, which are the thermal energy carried by the gas that crosses the slab with a thickness of 100 pc at a height of 1 kpc per year. {\it Lower right}: Energy loading factors for all runs as a function of time.}
    \label{fig:Outflows}
\end{figure*}

Figure~\ref{fig:Outflows} shows the mass (top) and energy (bottom) outflow rates (left) and loading factors (right). The yellow lines represent the model \SNnoFUVML, and the blue lines represent the model \SNnoFUVDT. The red dashed lines are the mean values obtained in the MFM simulation \citep{Steinwandel+23} as in their WLM-fid simulation. We measure these flow rates and loading factors at two different heights, at 1 kpc (solid) and 10 kpc (dashed).

One important feature is that while mass outflow and loading factor at 10 kpc are sometimes smaller than those at 1 kpc by as much as one dex, for the energy outflow rates and loading factors, there is only a factor of about 2 differences between 1 kpc and 10 kpc. This indicates that the hot wind transports most of the energy while the cool wind transports most of the mass, which is in very good agreement with other studies of multiphase galactic outflows.  
We find that such an important feature, as well as other trends, is reproduced in \SNnoFUVML. 

\subsection{Transition of Sub-sonic-Supersonic in the Outflows}
\label{sec:transition}

\begin{figure*}[ht!]
    \includegraphics[width=\textwidth]{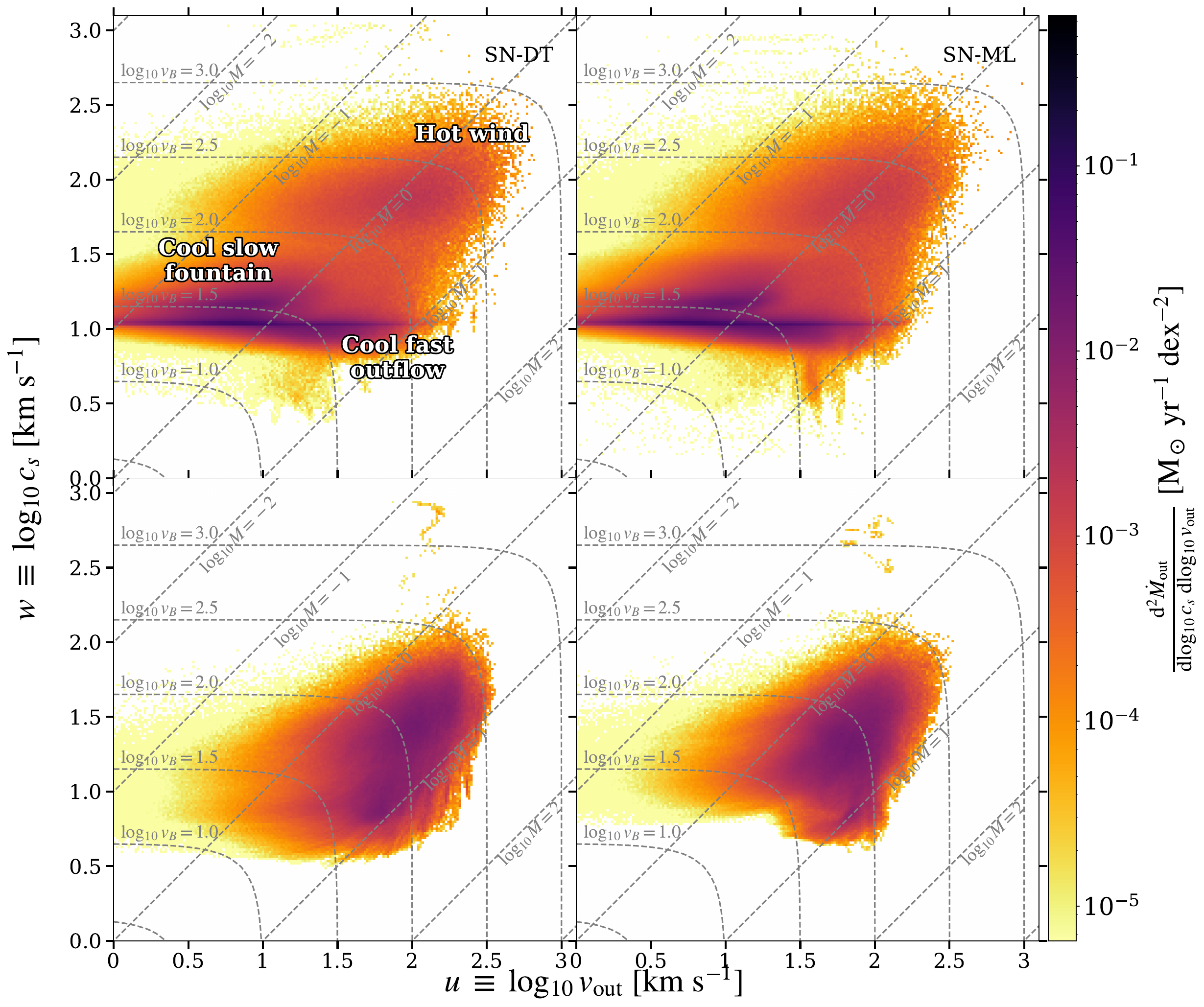}
    \caption{Two-dimensional joint PDFs of soundspeed and outflow velocity weighted by the mass outflow rate (flux) at height $z=1$ kpc (top) and at height $z=10$ kpc (bottom) for the models \SNnoFUVDT~(left) and \SNnoFUVML~(right). The mass outflow rates are averaged between 200 Myr with an interval of 2 Myr. The gray dotted lines in the pdfs define contours equal to Bernoulli velocity defined by Equation (\ref{eq:Bernoulli}).}
    \label{fig:phase_space_mass}
\end{figure*}
\begin{figure*}[ht!]
    \includegraphics[width=\textwidth]{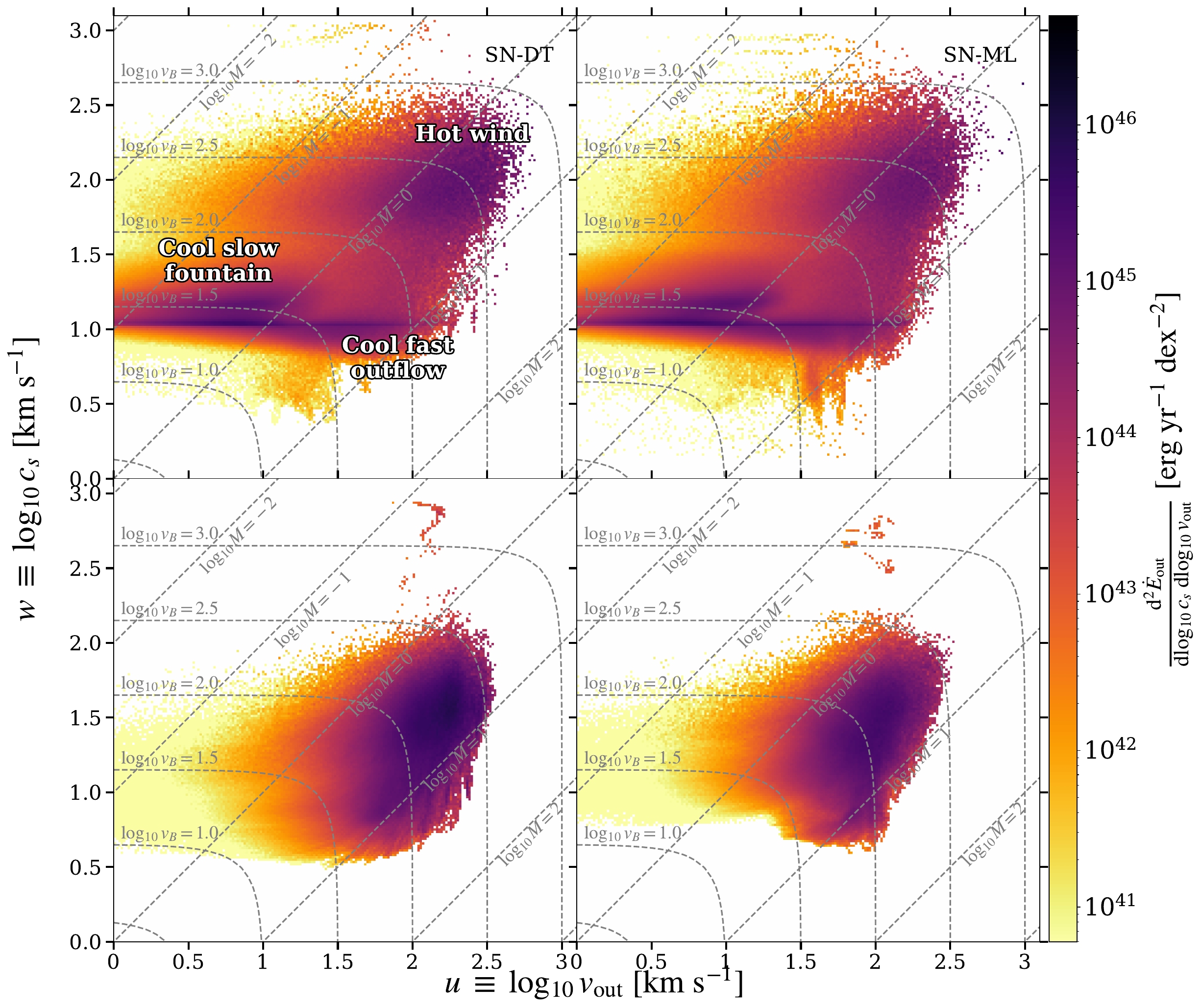}
    \caption{Two-dimensional joint PDFs of sound speed and outflow velocity weighted by the energy outflow rate (flux) at height $z=1$ kpc (top) and at height $z=10$ kpc (bottom) for the models \SNnoFUVDT~(left) and \SNnoFUVML~(right).}
    \label{fig:phase_space_energy}
\end{figure*}
Figure~\ref{fig:phase_space_mass} shows 2D joint probability distribution functions (PDFs) of sound speed $c_\mathrm{s}$ and outflow velocity $v_\mathrm{out}$ weighted by the mass outflow rate (flux), and Figure~\ref{fig:phase_space_energy} shows the same 2D PDFs weighted by energy outflow rate (flux). The top row measures these at a height of 1 kpc, while the bottom row measures these at 10 kpc. The left-hand side of the plots shows the results for \SNnoFUVDT, and the right-hand side shows those for \SNnoFUVML. We note that they generally follow the expected trend of a fully developed phase structure at low altitude (1 kpc), where the outflow can be subdivided into three parts, a cool slow outflow (the fountain with $c_\mathrm{s} \sim 10$ km s$^{-1}$ and $v_\mathrm{out} \lesssim 10$ km s$^{-1}$), the cool fast outflow (the entrained cold material in the hot wind, with $c_\mathrm{s} \sim 10$ km s$^{-1}$  and $v_\mathrm{out} > 30$ km s$^{-1}$) and the hot wind with $c_\mathrm{s} \sim 100$ km s$^{-1}$ and $v_\mathrm{out} \sim 100$ km s$^{-1}$. In this picture, the cool parts of the outflow carry most of the mass, while the hot parts of the outflow carry most of its energy. At higher altitudes, we find that the cool, slow part of the wind (i.e., the fountain) drops out in both cases, while the cool, fast outflow is mixed into the hot wind. The hot wind itself cools via adiabatic expansion and by mixing in the cool fast part of the wind, which reduces the average sound speed of the hot wind from  $c_\mathrm{s} \sim 100$ km s$^{-1}$ at 1 kpc height to about $c_\mathrm{s} \sim 30$ km s$^{-1}$ at 10 kpc height.
With these PDFs, it becomes clear why the mass outflow rate and mass loading in the time evolution drop by a dex in Figure~\ref{fig:Outflows} while the energy outflow rate and energy loading drop by a factor of two when we compare the results at 1 kpc and 10 kpc. We find that the hot wind carries most of the energy while there is almost no mass transport in the hot wind.
Since the bulk of the mass transport at 1 kpc is in the fountain, which carries very little energy, this explains the increase in the offset of mass and energy outflow (loading factors) at 10 kpc compared to the results obtained at 1 kpc. 

These results are in general agreement with similar studies such as \citet{Steinwandel+24a}, \citet{Kim+20} and \citet{Rey+24}. However, in our model, the hot wind at low altitude appears to be trans-sonic to mildly super-sonic, while in the other mentioned studies, the hot wind is mildly sub-sonic at 1 kpc. At higher altitudes, the hot wind transits, in agreement with these earlier studies, to the strongly super-sonic regime with Mach numbers of up to 30. This is happening in both models \SNnoFUVDT~and \SNnoFUVML. The reason for this shift in the sonic point of the wind as a function of altitude is likely related to the details in the cooling and heating physics implemented in each code. A detailed study of the physical conditions of the crossing in the sonic point as a function of altitude between different resolved feedback treatments is beyond the scope of this work.

Finally, we note that the phase structure in mass and energy at 10 kpc somewhat differs between the runs \SNnoFUVDT~and \SNnoFUVML. In \SNnoFUVDT, we find a similar phase structure as reported in \citet{Steinwandel+24a}. However, in \SNnoFUVML, we observe distinct tails extending to lower outflow velocities and sound speeds in the hot outflows, indicating that these outflows carry less momentum overall. This is an indication of the fact that the outflows for \SNnoFUVML~carry less momentum overall. This suggests that the hot outflows in \SNnoFUVML~are slower, causing them to stall at a smaller distance from the galactic midplane effectively. The outflow momentum at 10 kpc in \SNnoFUVML~is roughly half of that in \SNnoFUVDT, further supporting this interpretation shown in Table \ref{tab:momentum_at_10kpc}. Nevertheless, because the fixed timestep $\Delta t = 2000$ years are sufficiently shorter than timescales of gravitational interaction (Section \ref{sec:framework}), SN feedback for sparse regions ($<1~\mathrm{cm^{-3}}$) (Section \ref{sec:rec_timestep}), and cooling in dense regions (Table \ref{tab:cooling_timescale}), this discrepancy at 10 kpc could be just due to a sampling effect, as the top contours (1 kpc) are based on an average of $\sim 2\times 10^5$ particles per snapshot, while the lower contours (10 kpc) can use only $\sim 2\times 10^2$ particles per snapshot.
\begin{table}
    \centering
        \caption{The fraction (number) of SPH elements denser than SF threshold, a hydrogen density of $10^2 \mathrm{cm^{-3}}$, whose cooling timescale is shorter than the integration timestep. In direct simulations, the timestep is Equation (\ref{eq:CFL}) for all particles, and then the smallest one is adopted. The timestep $\Delta t$ for \SNnoFUVML~is fixed to 2000 yr. The numbers are averaged using the snapshots every 20 Myr over 300 Myr.}
    \begin{tabular}{cr}
    \hline\hline
         \SNnoFUVDT &  $4.0 \times 10^{-3}$ \% (1) \\
         \SNnoFUVML     &  0.50 \% (111)  \\
         \hline
    \end{tabular}
    \label{tab:cooling_timescale}
\end{table}

\begin{table}
    \centering
    
    \caption{Momentum of outflows within a slab with a thickness $\Delta z=1$ kpc at $z=10$ kpc. The momentum is averaged between 500 Myr.}
    \label{tab:momentum_at_10kpc}
    \begin{tabular}{ccc}
    \hline\hline
          & Momentum ($\mathrm{M_\odot ~kpc\>yr}^{-1}$) & mean $v_z (\mathrm{km \>s}^{-1})$ \\ \hline
         \SNnoFUVDT &  $3.6 \times 10^{-3}$ & $1.0 \times 10^2 $ \\
         \SNnoFUVML     &  $2.5 \times 10^{-3}$ & $7.7\times 10$ \\
         \hline
    \end{tabular}
\end{table}

\subsection{Gas Density Distribution for SN Explosions}
\label{sec:density_dist}
\begin{figure}[ht!]
    \includegraphics[width=0.45\textwidth]{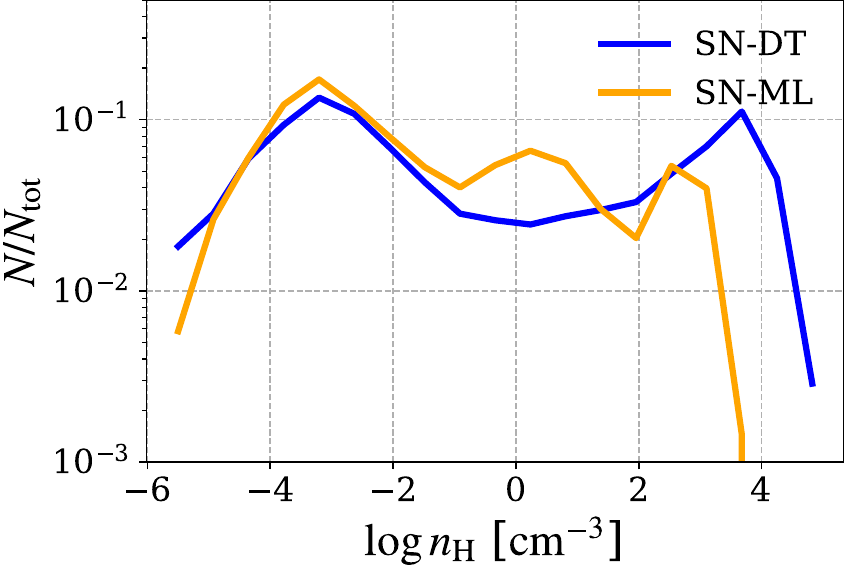}
    \caption{Probability distribution functions (PDF) of SNe as a function of the ambient gas density for 400 Myr listed in Table \ref{tab:models}.}
    \label{fig:SN_dist}
\end{figure}

Figure~\ref{fig:SN_dist} shows the probability distribution function of SNe as a function of the ambient density for the model \SNnoFUVDT~(blue) and the model \SNnoFUVML~(yellow). Each physical model shows a bimodal distribution that peaks at a number density of $10^{-3}$ and $10^3$ $\mathrm{cm^{-3}}$, which is consistent with previous results of the SN-environmental density distribution of \citet{Hu+2017}, \citet{Gutcke+2021} and \citet{Smith+21}.
The first peak around $10^{-3}~\mathrm{cm^{-3}}$ corresponds to the densities of typical super bubble interiors in those studies.
The second peak around $10^3 ~\mathrm{cm^{-3}}$ is larger than the SF threshold and is indicative of the SN-environmental distribution without early stellar feedback as outlined by \citet{Hu+2017} and \citet{Smith+21}.  
We note that this peak at high SN-environmental densities is shifted to lower environmental densities by half a dex in the model \SNnoFUVML~compared to the model \SNnoFUVDT. This fact is indicative that SN-feedback remains ineffective in those dense environments in the resolved simulation \SNnoFUVDT, and the ML algorithm in the simulation \SNnoFUVML~can, in fact, circumvent this behavior.

\subsection{Speed-up} \label{sec:speedup}
\begin{figure}[ht!]
\plotone{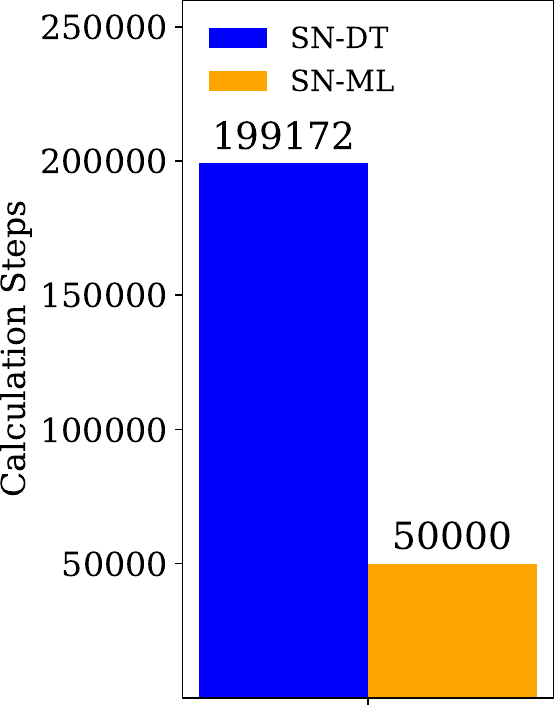}
\caption{The comparison of calculation steps between our conventional and our new simulation code accelerated by our surrogate modeling for 100 Myr. Our method can compress the calculation steps by a factor of four.
\label{fig:calcSteps}}
\end{figure}

Figure \ref{fig:calcSteps} shows the calculation steps required for running simulations with each physical condition listed in Table \ref{tab:models} for 100 Myr. For each case, applying our new ML model to the galaxy simulation code can be accelerated by a factor of four.
On ML runs in this paper, we chose a global timestep of 2000 yrs for integration and a prediction interval of 0.1 Myr. This setting grants our surrogate model a longer time for the inference than the one that requires 50 calculations to be completed. Thus, our surrogate modeling framework will not be a new bottleneck in the simulation code.
Additionally, the inference of ML in our framework is optimized for two main architectures, which are \textsc{ONNX} \citep{onnxruntime} for x86 architecture and \textsc{SoftNeuro} \citep{Hilaga+2021} for ARM architecture.
This will run the inference fast enough if a ML model with a larger number of parameters needs to be implemented in the future.

\section{Discussion}
\label{sec:discussion}

In this section, we discuss the results and put them in the broader context of similar feedback prescriptions currently used in the literature.

\subsection{Clustering of SNe}

It has recently been pointed out that the clustering of SN-explosions plays a major role in the potential of a (dwarf) galaxy to drive an outflow.
While there is still an ongoing debate on whether clustering \citep[e.g.,][]{Emerick+2018, Gutcke+2021, Smith+21, Hislop+2022,Hu+23, Steinwandel+23a} vs. SN-environmental density \citep[e.g.,][]{Hu+2016, Hu+2017, Steinwandel+2020, Steinwandel+24a, Steinwandel+24b}  is the primary driver of galactic outflows, there seems to be some consensus that they are ``two sides of the same coin''. For instance, \citet{Hu+2016, Hu+2017}, as well as the semi-analytical model, outlined that the lower environmental densities of SN-explosions are needed to generate a hot volume occupation fraction successfully launching the outflow. This intuition at the time has been built from the simulations of \citet{Hu+2017}, who, for the first time, invoked early stellar feedback which successfully quenched star formation on the cloud scale leading to lower star formation rates and thus lower outflow rates (but also loading factors) in comparison to simulations where SNe are the only feedback mechanism. While it is true that, generally speaking, the early stellar feedback lowers the environmental density, it also de-clusters star formation \citep[e.g.,][]{Smith+21, Hu+23}, which is considered to be the dominant factor. Theoretical models such as the one outlined in \citet{Naab+17} and \citet{Steinwandel+2020} make the argument that in order to establish a dominant volume filling phase one needs to have repeated explosion in gas with $n_\mathrm{H} <$ 0.1 cm$^{-3}$, which is coincidentally the environmental density range that is reported with early stellar feedback in simulations such as \citet{Hu+2017} and \citet{Smith+21}. However, they ignore that these conditions can be generated in a strongly clustered (SN-only) perspective, which produces clearly stronger outflows based on numerical simulations.

In Figure~\ref{fig:SN_dist}, we have seen that the density distribution of SN-environmental densities in direct simulations (blue, \SNnoFUVDT) shows two peaks around a number density of $10^{-3}$ and $10^3$, while the ML case (yellow, \SNnoFUVML) has only one prominent peak around a number density of $10^{-3}$.
However, since both models \SNnoFUVDT~and \SNnoFUVML~exhibit quite similar density distributions, as shown in Figure~\ref{fig:PhaseSpace}, it is not unreasonable to assume that a much greater number of massive stars form and explode in dense clouds in \SNnoFUVDT~than \SNnoFUVML.
The timestep is sufficiently short to resolve the cooling timescale even in dense regions, indicating that this issue is not due to a lack of time resolution as shown in Table \ref{tab:cooling_timescale}.

One possibility is that the simulation \SNnoFUVDT, with its pure thermal feedback, may be subject to overcooling in such dense regions, suppressing the expansion of SN shells immediately after the explosion. Since the simulation \SNnoFUVML~is bypassing this step with ML, one would never encounter such behavior leading to one prominent peak in explosions around 10$^{-3}$ cm$^{-3}$. There is some evidence for this since the total amount of SNe is around 13 percent for the model \SNnoFUVDT~and about 20 percent for the model \SNnoFUVML. Hence, subsequent SNe in the model \SNnoFUVML~can (somewhat by construction) occur in lower-density environments.
To what degree the ML framework helps to model more accurate thermal energy and momentum input in high-density regions in the simulation \SNnoFUVML~remains somewhat hard to say as we find an undershoot of momentum in low-resolution simulations of isolated bubbles (see Appendix. \ref{sec:fidelity}). It seems that the diffuse bubble can expand in high-density star-forming regions more efficiently, and as a result, short-lived stars may tend to explode in regions with much smaller densities than the SF threshold.

It is also interesting to compare \SNnoFUVDT~and \SNnoFUVML~with similar models of a star formation efficiency of 2 percent per free-fall time without photo-ionization: \texttt{noPE-noPI-SN} in \cite{Hu+2017}, \texttt{FIXED\_SN\_ENERGY} in \cite{Gutcke+2021}, and \texttt{SFE02noPI} in \cite{Hislop+2022}. Their initial condition of the dwarf galaxy is similar settings as \cite{Steinwandel+23}. In the implementation of SN feedback in \cite{Hislop+2022}, the thermal energy of SNe was injected at the ambient number density of $n_\mathrm{H} < 10 \>\mathrm{cm^{-3}}$ where individual SNe are well resolved and the expected outer momentum is input in dense regions \citep{Hu+2016,Steinwandel+2020}.
The ambient density distributions of SN explosions for \texttt{SFE02noPI} also have a bimodal distribution with peaks around $10^{-4}$ and $10^0 ~\mathrm{cm^{-3}}$.

\SNnoFUVML~gains more SNe around $10^0 ~\mathrm{cm^{-3}}$ compared to \SNnoFUVDT.
This can be because our ML model projects the prediction of the high resolution (1 $\mathrm{M_\odot}$) to a simulation with low resolution (4 $\mathrm{M_\odot}$).
The large population of SNe explodes in the evacuated regions of previous SNe, which makes the feedback more efficient.
After the first SN explodes on a gas cloud in our resolved simulations with a mass resolution of 1 $\mathrm{M_\odot}$, a diffused bubble can reach a density of $10^{-2} - 10^{-4}$ $\mathrm{cm^{-3}}$, close values reported in \cite{Gutcke+2021} and \citet{Hislop+2022}.

\subsection{Phase Structures of the Wind}

The studies most relevant to this work are \citet{Fielding+18}, \citet{Kim+20}, \cite{Steinwandel+23}, and \citet{Rey+24} as they utilize comparable models for handling the ISM in dwarf galaxies. 
Our results agree with the previous simulations that the cool wind caries most of the mass while the hot wind carries most of the energy, as noted in Sec.~\ref{sec:transition}. The features in the 2D joint PDF of sound speed and outflow velocity weighted by the mass outflow rate (flux) in Figure~\ref{fig:phase_space_mass} and energy outflow rate (flux) in Figure~\ref{fig:phase_space_energy} are consistent with those of previous works. 
For winds at low altitude, \citet{Fielding+18} and \cite{Kim+20} found that cool outflow ($T< 10^4$ K i.e., $\log_{10}{\left[c_s/\left(\mathrm{km~s}^{-1}\right)\right]}<1.18$) carries most of the mass while the hot outflow ($T>10^5$ K i.e., $\log_{10}{\left[c_s/\left(\mathrm{km~s}^{-1}\right)\right]}>1.68$) carries most of the energy at $z<1 ~\mathrm{kpc}$.
\cite{Rey+24} and \cite{Steinwandel+23}, using a similar simulation setup to this study, also showed a sub-to-supersonic transition in the wind at 1 and 3 kpc.

Literature on the simulations of even heavier galaxies also show the transition between sub-sonic wind at low altitude ($z\sim1$ kpc) and super-sonic wind at high altitude ($z>10$ kpc) in 2D PDFs likely caused by the fountain.
\cite{Steinwandel+24a, Steinwandel+24b}, which used an isolated galaxy with ten times heavier virial mass than this paper, showed the sub-to-super sonic transition of hot winds between 1 kpc and 10 kpc.
\citet{Pandya+2021} using one hundred times heavier galaxy than this paper, showed the hot super-sonic wind at $\sim 25$ kpc.
The majority of the hot wind at the high altitude is supersonic, while the subsonic gas drops off.

Lastly, we note that their simulations include additional stellar feedback, such as photoionization and FUV heating from individual stars, which may be crucial for heating the ISM. However, since the transition of phase structures in the wind aligns with our results, this suggests that the hot gas produced by SN-feedback plays a significant role in driving the super-sonic winds that can reach high altitudes.

\section{Conclusion and Outlook}
\label{sec:conclusions}

We have presented new high-resolution simulations of an isolated dwarf galaxy with our ML model for resolved SN feedback.
The key findings of our study can be summarized as follows:
\begin{enumerate}
    \item Our new framework \textsc{ASURA-FDPS-ML} for star-by-star galaxy simulations using a surrogate model for SN feedback shows the capability to reconstruct numerical direct simulations' results in morphological structure, star formation history, and outflows. 
    
    \item We find that the run of an isolated dwarf galaxy with our ML framework achieved a speedup by a factor of four.

    \item Regarding multi-scale gas structures, the phase space diagrams and multi-phase outflows resemble between \SNnoFUVDT~and \SNnoFUVML. The mass and energy outflow rates and loading factors also have agreements.
    
    \item As expected from previous studies, the hot wind carries most of the thermal energy in the outflow in our simulations. Most hot super-sonic gas can reach high altitudes despite the gravitational potential from the host galaxy.
    
    \item We also found a few discrepancies. 
    The number of SNe in the dense regions decreases when we use ML. This could imply SN feedback in dense regions is more effective in \SNnoFUVML~because it projects predictions with a mass resolution of 1 $\mathrm{M_\odot}$ to simulations with a mass resolution of 4 $\mathrm{M_\odot}$.
    We found \SNnoFUVML~have a bit low momentum particles at $z=10$ kpc. That could imply the prediction has difficulties predicting the momentum of hot particles, which will be addressed in future works.
\end{enumerate}

We show an isolated dwarf galaxy simulation with a surrogate model and showcase the first practical usage and substantial speedup. We expect, with this new framework, star-by-star simulations of heavier galaxies such as LMC-size and MW-size with a mass of $10^{11}$ to $10^{12} ~\mathrm{M_\odot}$ and a mass resolution of 1 to $10 ~\mathrm{M_\odot}$, challenging for even state-of-the-art numerical simulations can be run with a reasonable runtime.

Nevertheless, for more detailed studies on star formation in a galactic environment using star-by-star simulations, we may also need to resolve individual HII regions due to photoionization, where the timescale for cooling/heating may become very short, especially in dense regions. In that case, one of the solutions would be to implement a surrogate model for photoionization in dense regions using such as a $\mathrm{Str\ddot{o}mgren}$ type approximation to localize the effect \citep[e.g.,][]{Hu+2017}.
That would make it possible also to study so-called triggered star formation \citep[e.g.,][]{Melioli+06, Nagakura+09, Hobbs+20} in the environment of galaxies.

While we have focused on the SNe remnants after Sedov phase, the problem of the short timesteps also appears in the earlier phase of free expansion. Such expansion is terminated when the swept-up mass becomes equal to the ejecta mass of the remnant of a few solar masses \citep[e.g.,][]{Reynolds2008}, and the reverse shock decouples from the forward shock, propagating backward into the vacuum generated by the forward shock. 
As the reverse shock heats the gas inside the cavity generated by the forward shock to temperatures of $\sim 10^7$ K,
the injection of the SN-energy has been generally handled as a single thermal injection event so far as a ``subgrid model." 
To resolve this phase more accurately, one would need to have at least 10 to 100 times higher mass resolution than the ejecta mass.
The kinetic injection schemes have been applied in many successful simulation endeavors of the ISM \citep[e.g.,][]{Hu+2016}, but this scheme does also require not only a high mass resolution but also a sufficient temporal resolution. The machine learning method similar to the one developed in this study may also prove effective in modeling these phases.

\section*{Acknowledgments}

The authors thank Carolina Cuesta-Lazaro, Chang-Goo Kim, Eve Ostriker, and Philip F. Hopkins for fruitful discussions.
This work was supported by JSPS KAKENHI Grant Numbers 22H01259, 22KJ0157, 25K01046, 21K03614, 22J23077, 22KJ1153, 23K03446, and 24K07095, 23K20035, 24H00004, and MEXT as ``Program for Promoting Research on the Supercomputer Fugaku'' (Structure and Evolution of the Universe Unraveled by Fusion of Simulation and AI; Grant Number JPMXP1020230406; Project ID: hp230204, hp240219).
KH is supported by JSPS Research Fellowship for Young Scientists, JEES $\cdot$ Mitsubishi corporation science technology student scholarship in 2022, The University of Tokyo doctoral fellowship with the IIW program, and the JSPS International Leading Research (ILR) program (JP22K21349), and thanks CCA at the Flatiron Institute for hospitality and the National Science Foundation under Cooperative Agreement PHY-2019786 (The NSF AI Institute for Artificial Intelligence and Fundamental Interactions, \url{http://iaifi.org/}) for support while a portion of this research was carried out. YH is supported by the grant OISE-1927130: The International Research Network for Nuclear Astrophysics (IReNA), awarded by the US National Science Foundation (NSF). UPS is supported by a Flatiron Research Fellowship (FRF) at the Flatiron Institutes Center for Computational Astrophysics (CCA). The Flatiron Institute is supported by the Simons Foundation.

%

\vspace{5mm}

\facilities{Popeye (Flatiron Institute), Cray XC50 (CfCA, NAOJ), Wisteria/BDEC-01 Aquarius (University of Tokyo), Fugaku (Riken Center for Computational Science)}

\software{Astrophysical Multi-purpose Software Environment\citep[\textsc{AMUSE};][]{Pelupessy+2013, Portegies+2013, Portegies+2018}, Cloudy \citep{Ferland+1998, Ferland+2013, Ferland+2017}, Open Neural Network Exchange \citep{onnxruntime}, SoftNeuro \citep{Hilaga+2021}, TensorFlow \citep{tensorflow}, PyTorch \citep{pytorch}, yt \citep{Turk+2011}, NumPy \citep{Numpy}, Matplotlib \citep{Matplotlib}}



\appendix

\section{Convergence Test of our SPH code} \label{sec:convergence}
\begin{figure*}[ht!]
    \plotone{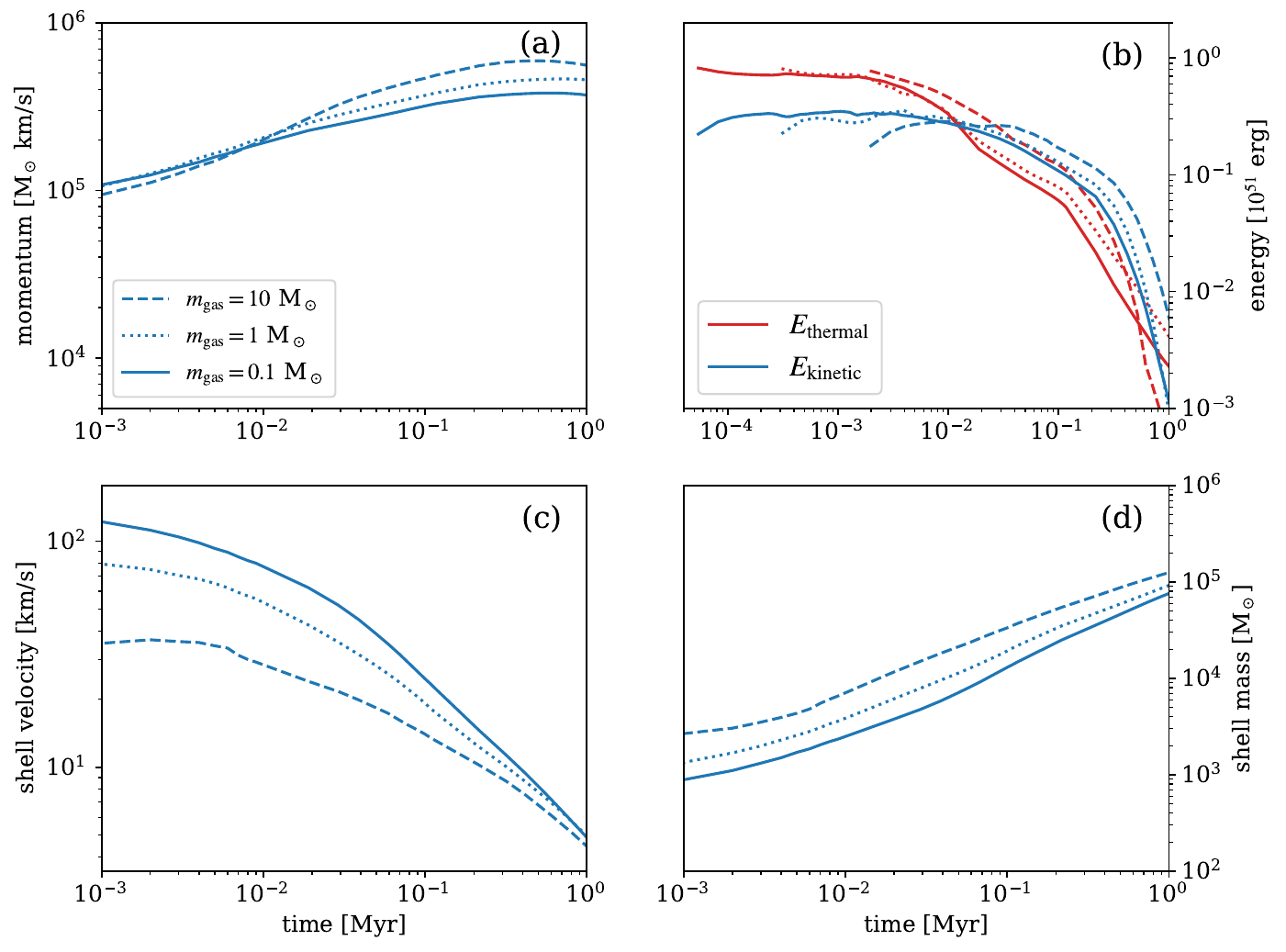}
    \caption{Evolution of a SN shell in the cold neutral medium (initial temperature $T = 100$ K and density $n_0 = 45$ ${\rm cm^{-3}}$) with a metallicity of 0.1 $\mathrm{Z_\odot}$, representing the ISM in the environment of a dwarf galaxy, as modeled following the convergence test in \citet{Hu+2016} and \citet{Hirashima+23a}. Three different mass resolutions are considered: $m_{\rm gas} = 0.1$ (solid line), 1 (dotted line), and 10 (dashed line) M$_\odot$. Panel (a): Linear momentum of the shell. Panel (b): Thermal energy (red) and kinetic energy (blue). Panel (c): Shell velocity, defined as the total momentum divided by the shell mass. Panel (d): Shell mass. The shell is defined as all particles with temperatures $T < 2 \times 10^4$ K and velocities $v > 0.1$ km s$^{-1}$.}
    \label{fig:conv_SPH}
\end{figure*}

The results of SPH simulations depend on the resolution. We, therefore, performed a numerical convergence test for single SN evolution using three different mass resolutions.
Figure \ref{fig:conv_SPH} shows the physical quantities of SN shell evolution, following a conversion test in \citet{Hu+2016} and \citet{Hirashima+23a}.
In this test, we performed the simulations for 1 Myr after injecting thermal energy of $10^{51}$ erg into 100 SPH particles.
As the background gas distribution, we adopted a molecular cloud with a mass of $10^6~\mathrm{M_\odot}$, an initial temperature of 100 K, and a metallicity of 0.1 $\mathrm{Z_\odot}$.
We conclude that the time evolution of physical quantities almost converges at 1 $\mathrm{M_\odot}$, which is the mass resolution used for our training dataset.
This mass resolution required for the convergence is consistent with \citet{Steinwandel+2020}, which explored the resolution convergence with broad physical conditions.

\section{Pre-evolved Initial Condition} 
\label{app:pre-evolution}

\begin{figure*}[ht!]
    \plotone{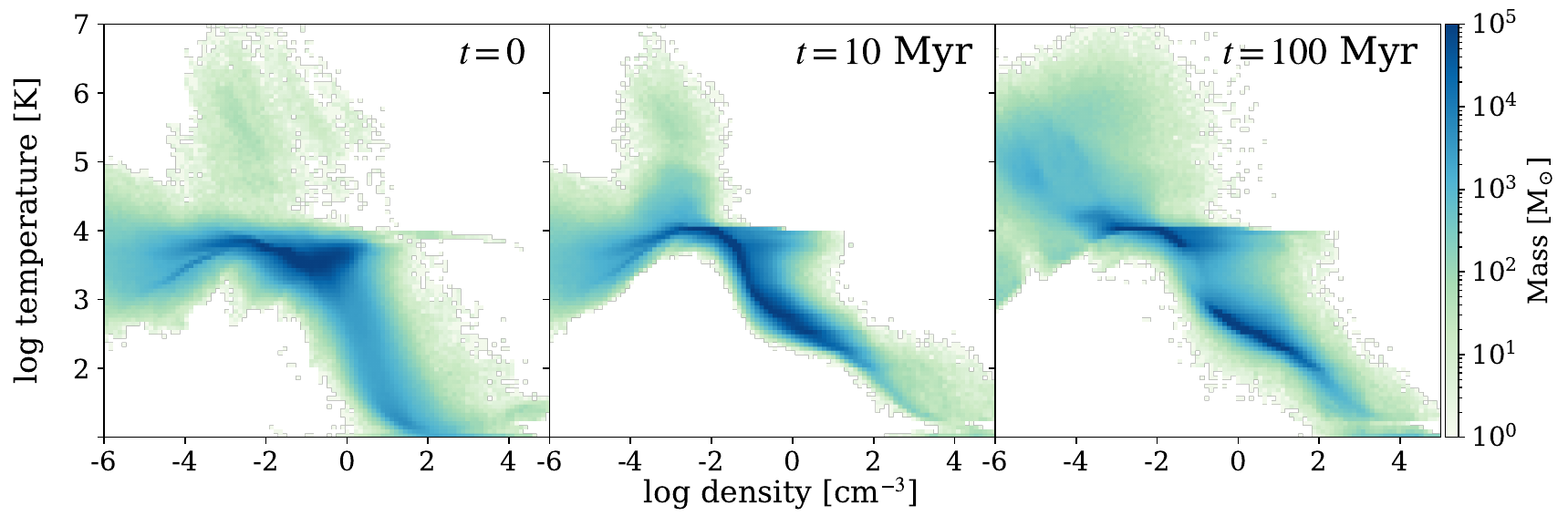}
    \caption{Phase structures of \SNnoFUVDT~({\it left}) as listed in Table \ref{tab:models} at $t=0$, $t=10$ Myr, and $t=100$ Myr. Density-temperature phase diagrams of mass.}
    \label{fig:PhaseSpace_comp}
\end{figure*}

While initial condition generators for disk-galaxy simulations \citep[e.g., ][]{Springel2005} can reconstruct summary statistics, they generally create smoothed gas distribution without detailed structures such as filaments with strong density contrasts and large bubbles created by SNe. The smoothed distribution could cause unrealistic starbursts. 
To avoid this issue, simulations typically require hundreds of megayears to develop turbulent fields and density contrasts driven by stellar feedback, which means early results must be discarded -- a significant computational waste.
In this study, to save computational costs, we instead use a snapshot from \citet{Steinwandel+23} that was generated using different codes and cooling functions.

During 500 Myr of pre-evolution, only a small amount of gas mass is converted into stellar mass; the resulting stellar and gas mass at 500 Myr are $2 \times 10^7 ~\mathrm{M_\odot}$ and $3.96 \times 10^7 ~\mathrm{M_\odot}$, respectively.
The gas mass change is only 0.67 percent, which is negligible for evolutionary comparisons.
While preserving the overall mass distribution of the initial galaxy disk, the pre-simulated galaxy exhibits large bubbles and small-scale density contrasts in the ISM that help prevent artificial starbursts.

For consistency, our simulations do not inherit stars born during the pre-simulation; all SNe in our simulations originate from stars formed within our simulation timeframe.
This treatment potentially impacts the ISM phase structure during the transition between numerical solvers. 
Figure \ref{fig:PhaseSpace_comp} compares ISM phase structures at $t=0$, 10 Myr, and 100 Myr, where the $t=0$ snapshot corresponds to the $t=500$ Myr snapshot in \citet{Steinwandel+23}.
Between $t=0$ and $t=10$ Myr, we find a reduction in gas with a temperature of $T<10^2$ K and a number density of $n>1 ~\mathrm{cm^{-3}}$. The gas kept missing even at $t=100$ Myr, which is likely due to changes in cooling functions.
The dense gas undergoes strong cooling in the early period of our simulation, resulting in an artificial starburst as shown in Figure \ref{fig:SFR}.
At $t=10$ Myr, fewer gas particles exhibit temperatures above $10^5$ K compared to $t=0$, since we did not incorporate stellar feedback from massive stars formed during pre-simulation.
These effects, however, diminish by $t=100$ Myr, after the artificial initial starburst subsides and SN feedback effects become prominent.

We conclude that the ISM phase structure is initially affected by the sudden change in the thermodynamic model and the temporary absence of stellar feedback. 
However, this deficiency in high-temperature gas is largely mitigated following the artificial initial starburst period, when SN feedback effects become prominent.

\section{Surrogate Modeling} \label{sec:surrogate_model}

We adopt U-Net, a model based on convolutional neural networks (CNNs). We first map SPH particles into voxels (uniform mesh in 3D) with interpolation, which is necessary to process the data with CNNs. After predicting the results, we sampled SPH particle data from the voxels. Such methods have been applied to hydrodynamical simulations \citep[][]{Hirashima+23a, Hirashima+23b, Chan+24} and $N$-body simulations \citep[][]{He+19, Jamieson+23, Jamieson+24, Legin+24}. In this section, we describe the interpolation method and our ML model in detail.

\subsection{Voxel Generation}
To train CNN-based models with the results from SPH simulations, we converted the physical quantities expressed with the gas particles to voxels with 3D Cartesian grids.
Specifically, the density, temperature, and 3D velocities are represented as five 3D scalar fields of size $64^3$ voxels with one side of 60 pc.
Following SPH method, a physical quantity $f_i$ of a voxel $i$ is interpolated using the SPH particles $j$ and kernel: 
\begin{equation}
    f_i = \sum _j m_j \frac{f_j}{\rho_j} W(r_{ij}, h),
    \label{eq:SPHconv}
\end{equation}
where $m_j$, $\rho_j$, $r_{ij}$, $h$, and $W(r_{ij}, h)$ are the mass, density, distance between the particles $i$ and $j$, and a smooth function.
The number of neighboring particles is set to 64. 
Practically, we also applied a normalized interpolation based on Sheperd's method \citep[inverse-distance-weighted interpolation;][]{Shepard+68,Price07} to improve the precision of quantities interpolated into voxels by mitigating effects related to particle distribution. The normalization is achieved by ensuring that the sum of the interpolation weights $w(i)$ is equal to unity.
The weight $w(i)$ of particle $i$ for Equation (\ref{eq:SPHconv}) is defined as the following:
\begin{align}
    w(i) &:= \sum _{j=1} ^{N} \frac{m_j}{\rho_j}  W(|\bm{r}_{i}-\bm{r}_{j}|, h_i) \nonumber\\
    &= \sum _{j=1} ^{N} \frac{m_j}{\rho_j}  W_{i,j}.
    \label{eq:weight}
\end{align}
The normalized quantities $\tilde{f}(i)$ (i.e., normalized temperature and velocities) can be obtained by dividing Equation (\ref{eq:SPHconv}) with Equation (\ref{eq:weight}) as
\begin{equation}
    \tilde{f}(i) := f(i) / w(i).
    \label{eq:normalization}
\end{equation}

Since the hydrodynamics of ISM is compressive, the physical quantities, such as density and temperature, vary in several magnitudes.
Thus, those quantities are normalized to enable the model to learn the dataset effectively. We adopt the normalization:
\begin{align}
\rho^* &:= \log_{10} \rho, \\
T^* &:= \log_{10} \tilde{T}.
\end{align}
Velocities have a bimodal distribution.
To improve the accuracy, the 3D velocities were distributed into six colors.
\begin{align}
    v_{\cdot,p}^* &:=
    \begin{cases} 
        \log_{10} \tilde{v}_{\cdot} & \text{if } \tilde{v}_{\cdot} > 0 \\
        0      & \text{if } \tilde{v}_{\cdot} \leq 0
    \end{cases}\\
    v_{\cdot,n}^* &:=
    \begin{cases} 
        0 & \text{if } \tilde{v}_{\cdot} \geq 0 \\
        \log_{10} (-\tilde{v}_{\cdot})  & \text{if } \tilde{v}_{\cdot} < 0
    \end{cases}
\end{align}
where $v_{\cdot} \in \left\{  v_x, v_y, v_z \right\}$.
Given by these conversion, these eight features, $\left \{ \rho^*, T^*, v_{x,p}^*, v_{x,n}^*, v_{y,p}^*, v_{y,n}^*, v_{z,p}^*, v_{z,n}^* \right \}$, are used in training.

\subsection{3D-UNet}
We used U-net \citep{Ronneberger+15}, but we extended the internal dimension from 2D to 3D. We adopted a channel size of eight, a batch size of one, and a patch size of one.
The mean squared error (MSE) is used for the loss function with an equal weight for each channel.
The loss is minimized with ADAM optimizer \citep{Kingma+2014} with a learning rate of $10^{-5}$.
The model $\mathcal{M}$ is trained to learn the relation between input $X$ and output $y$, where $X$ and $y$ represent the distribution of physical quantities before and after 0.1 Myr of the SN explosion, respectively.
Suppose the trainable parameter $\theta$, the predicted distribution $\hat{y}$ is written as the following:
\begin{equation}
    \hat{y} = \mathcal{M}(X \mid \theta).
\end{equation}

\subsection{Sampling}
We perform Gibbs sampling, a Markov chain Monte Carlo method, to extract gas elements for SPH simulations from the predicted voxel data. Gibbs sampling generates an approximate sequence of samples by iteratively using the conditional probabilities of each variable.
Using the predicted distribution of density $\hat{\rho}$, the probability is written as
\begin{equation}
    p(x,y,z) :=  \frac{\hat{\rho}(x,y,z)}{\sum_x \sum_y \sum_z \hat{\rho}(x,y,z)},
\end{equation}
as the joint probability where $x, y, z$ are the positions of voxels. We added small perturbations to the particle positions to avoid overlap, which is a physically unrealistic situation. 
The perturbation values are drawn from a uniform distribution within the range [0, $\Delta w$], where $\Delta w$ corresponds to the width of one voxel. For our SN model, $\Delta w$ is approximately 1 pc, which is calculated as $\Delta w = 60 ~\mathrm{pc} ~/ 64 \sim 1$ pc.
The burn-in period is set to be 1000.
For a given joint distribution $p(x, y, z)$, the Gibbs sampling generates particles as follows:
\begin{enumerate}
    \item Choose an initial state $(x^{(0)}, y^{(0)}, z^{(0)})$.
    \item For each iteration $t = 1, 2, \ldots, N$:
    \begin{enumerate}
        \item Sample $x^{(t)}$ from $p\left(x \mid y^{(t-1)}, z^{(t-1)}\right)$.
        \item Sample $y^{(t)}$ from $p\left(y \mid x^{(t)}, z^{(t-1)}\right)$.
        \item Sample $z^{(t)}$ from $p\left(z \mid x^{(t)}, y^{(t)}\right)$.
    \end{enumerate}
\end{enumerate}
The mass is conserved before and after the prediction by setting $N$ as the number of particles in the box. The other physical quantities of sampled particles, such as temperature and 3D velocities, are assigned based on the values of the corresponding voxel at the same coordinates.

\begin{figure}[t!]
\vspace{11pt}
    \plotone{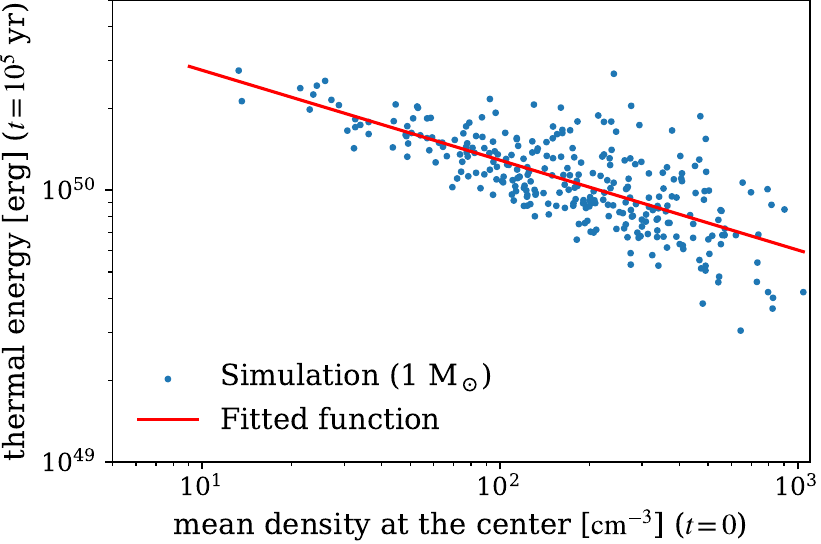}
    \caption{Relationship between the mean central density before SN and the thermal energy of particles hotter than $10^4$ K at 0.1 Myr after the SN explosion. The linear function is fitted to 300 training datasets for the 3D-UNet.
    \label{fig:relation}}
\end{figure}
Reproducing the thermal energy and outward momentum is crucial for modeling SN feedback. However, since only a small fraction of particles in the voxels have high temperatures and large momentum, standard sampling methods like Gibbs sampling may face difficulties in sampling these particles. Therefore, by referencing the predicted total thermal energy at 0.1 Myr, we ensure sampling of a sufficient number of gas particles with high temperature ($>10^4$ K). This threshold is determined using a linear function fitted to the training data, relating the initial mean density within a sphere of 5 pc radius at $t=0$ to the final thermal energy of particles hotter than $10^4$ K at $t=0.1$ Myr shown in Figure \ref{fig:relation}.
After Gibbs sampling, the thermal energy of the reconstructions is compared to the estimated thermal energy to determine if sampling should continue. Sampling is repeated until the difference becomes within a threshold. In this paper, the threshold is 5 percent, and the iteration occurs only a few times.

\subsection{Fidelity}
\label{sec:fidelity}

\begin{figure*}[ht!]
    \plotone{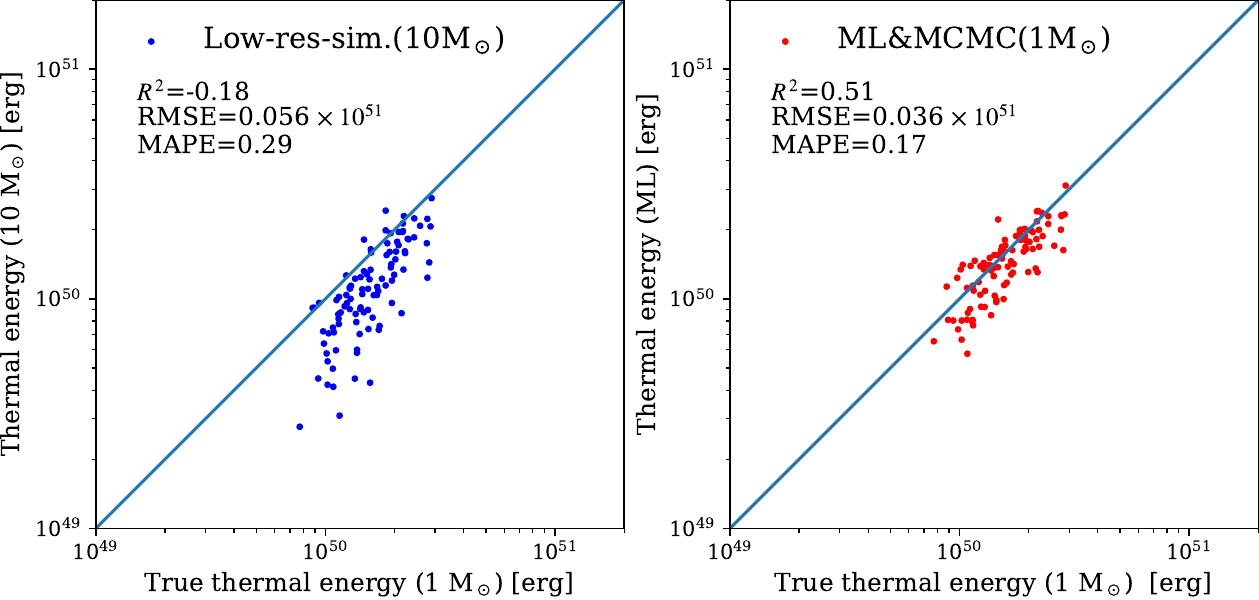}
    \caption{Fidelity evaluation in thermal energy. Using the high-resolution simulation (1 M$_\odot$ resolution; $x$-axis) results as a baseline, we compared our method ($y$-axis, {\it right}) with the corresponding low-resolution simulation (10 M$_\odot$ resolution; $y$-axis, {\it left}). We evaluate 100 test data by the determination coefficient $R^2$, root mean squared error (RMSE), and mean absolute percentage error (MAPE).}
    \label{fig:thermal_comparison}
\end{figure*}

\begin{figure*}[ht!]
    \plotone{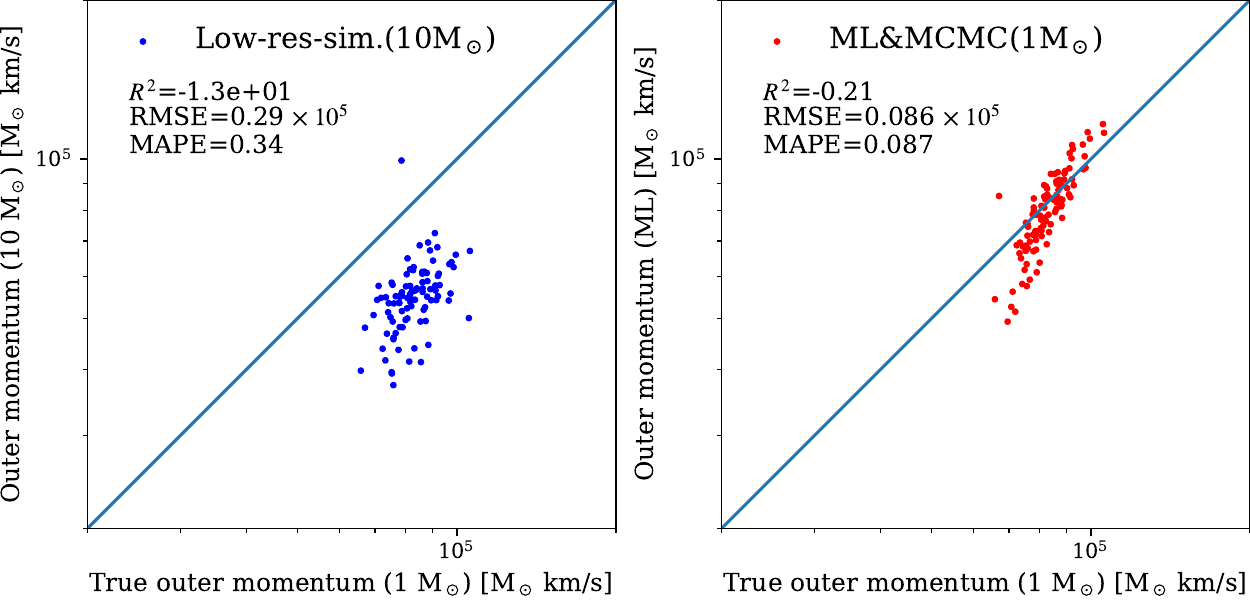}
    \caption{Fidelity evaluation in radial outward momentum. Other settings are the same as Figure \ref{fig:thermal_comparison}.}
    \label{fig:mom_comparison}
\end{figure*}

We evaluated the fidelity of our surrogate model including the prediction and sampling together. 
Comparison between the direct simulation and surrogate model is not simple. Since a cooling model is included, the thermal energy is lost during the evolution, and therefore, the energy is not conserved even in the full direct simulations. The sampling of the particles also produces some scatter in the total energy. 
Therefore, we performed lower resolution simulations (10 M$_\odot$) and compared the scatter to the surrogate model. 
The low-resolution simulations (10 M$_\odot$) were performed using initial conditions with the same turbulence field as that for the high-resolution ones (1 M$_\odot$). 
In Figure \ref{fig:thermal_comparison}, we present the comparison of the thermal energy. The discrepancy is evaluated with the determination coefficient $R^2$, root mean square error (RMSE), and mean absolute percentage error (MAPE). Although both the low-resolution and surrogate models do not perfectly match the results of the high-resolution simulations, the surrogate models scatter around the linear relation. On the other hand, the low-resolution simulations systematically lose the thermal energy. 
We also compared the radial outward momentum (see Figure \ref{fig:mom_comparison}). Similar to the thermal energy, low-resolution simulations systematically underestimate the momentum. Thus, our surrogate model performed better than the low-resolution simulations.


\bibliography{surrogate}{}
\bibliographystyle{aasjournal}



\end{CJK*}
\end{document}